\documentclass[]{amsart}
\usepackage{amsmath,amssymb}

\usepackage{CJK}
\usepackage[colorlinks=true, linkcolor=cyan, citecolor=blue, urlcolor =black]{hyperref}
\usepackage{amsmath,mathtools}
\usepackage{amsthm}
\usepackage{bbold}
\usepackage[utf8]{inputenc} 
\usepackage{graphicx}
\usepackage{braket}
\usepackage{ragged2e}
\usepackage{wrapfig}
\usepackage{color}
\usepackage{pifont}
\usepackage{siunitx}
\usepackage[english]{babel}
\newtheorem{theorem}{Theorem}

\usepackage[colorinlistoftodos]{todonotes}

\newcommand{\circledone}[1]{\large \ding{192} \normalsize}
\newcommand{\circledtwo}[1]{\large \ding{193} \normalsize}

\newcommand{\TT}[0]{\mathcal{T}}

\newcommand{\II}[0]{\mathcal{I}}

\newcommand{\GG}[0]{\mathcal{G}}

\newcommand{\XX}[0]{\bf{X}}
\newcommand{\ZZ}[0]{\mathbb{Z}}
\newcommand{\RR}[0]{\mathbb{R}}

\newcommand{\ad}[0]{\bar{\Delta}}
\newcommand{\C}[0]{\mbox{cap}}
\newcommand{\ff}[0]{h}

\newcommand{\tubeF}{2\cos \left( \frac{2 \pi}{10} \right)}
\newcommand{\tubeFsquared}{ \left(  2\cos \left( \frac{2 \pi}{10} \right)    \right)^2}

\usepackage[utf8]{inputenc}
\usepackage[english]{babel}

\newtheorem*{ack}{Acknowledgements}

\newtheorem{proposition}{Proposition}[]
\newtheorem{manualtheorem}{Theorem}[]

\usepackage[all,cmtip]{xy} 

\usepackage{xcolor}

\begin{document}

\title{\sffamily Gap Sets for the Spectra of Cubic Graphs} 

\author{Alicia J. Koll\'{a}r}
\address{Department of Physics and JQI, University of Maryland, College Park, MD 20742, USA}
\author{Peter Sarnak}
\address{Department of Mathematics, Princeton University, Princeton, NJ 08540, USA}


\date{\today}

\begin{abstract}
We study gaps in the spectra of the adjacency matrices of large finite cubic graphs. It is known that the gap intervals $(2 \sqrt{2},3)$ and $[-3,-2)$ achieved in cubic Ramanujan graphs and line graphs are maximal. We give constraints on spectra in [-3,3] which are maximally gapped and construct examples which achieve these bounds. 
These graphs yield new instances of maximally gapped intervals. We also show that every point in $[-3,3)$ can be gapped by planar cubic graphs.  Our results show that the study of spectra of cubic, and even planar cubic, graphs is subtle and very rich.
\end{abstract}
\maketitle 

\section{Introduction}\label{sec:intro}
By a cubic graph we mean a finite $3$-regular connected graph with no loops or multiple edges. Denote the set of such graphs by $\bf{X}$ and the subset of $\bf{X}$ which can be realized as planar graphs by $\bf{X}_{Planar}$. 
For $Y \in \bf{X}$ we denote the adjacency matrix of $Y$ by $\ad_Y$ to highlight its equivalence to the graph Laplacian.
The spectrum of $\ad_Y$, denoted by $\sigma(Y)$ is contained in $[-3,3]$ and contains $3$ (as a simple eigenvalue). 
The problem of constructing large $Y$'s with gaps in their spectra arises in different contexts. In combinatorics and engineering applications, a gap at $3$ defines ``cubic expanders'', an apparently very fruitful structure \cite{H-L-W}. In our recent work  \cite{K-F-S-H}
 on microwave coplanar waveguide resonators it is the gap at the bottom $-3$ that is critical. In chemistry the stability properties of carbon fullerene molecules are dictated by the gap at $0$ for the case of closed shells \cite{F-M}.
 Our goal is to determine what gaps can be achieved by large elements of $\bf{X}$ and $\bf{X}_{planar}$, and in particular to identify maximal gap intervals and sets.

To formulate our results, we make some definitions. A closed subset $K$ of $[-3,3]$ is spectral (resp planar spectral) if there are arbitrarily large, or equivalently infinitely many, $Y$'s in $\bf{X}$ (resp $\bf{X}_{Planar}$) with $\sigma(Y) \subset K$. The complement in $[-3,3]$ of a spectral set is called a gap set. A closed superset of spectral sets is spectral, and we seek minimal spectral sets, or equivalently, maximal gap sets.

The first question is whether every point $\xi$ in $[-3,3)$ can be gapped, meaning that $\xi$ is contained in an open neighborhood $U_\xi$ which is a gap set. One of our main results answers this:

\begin{theorem}[ ]
\label{thm:planargaps}
 Every point in $[-3,3)$ is planar gapped.
\end{theorem}

Fekete's theorem \cite{Fe} gives a lower bound of $1$ for the transfinite diameter, or equivalently the capacity, of a closed subset of $\mathbb{C}$ which contains infinitely many algebraic integers as well as their conjugates. For definitions and properties of capacity see \cite{Ah}. We apply this together with combinatorial arguments to give general lower bounds for the size and shape of a spectral set $K$. Remarkably, these bounds are sharp, in that they are achieved for certain $K$'s.

\begin{theorem}\label{thm:stronggap}
Let $K$ be a spectral set, then 
\begin{enumerate}
\item[(i)]  $\C(K) \geq 1$.

\item[(ii)] If $I$ is an interval contained in $(-1-\sqrt{2}, 1 + \sqrt{2})$ whose length is greater than $2$, then 
$$ I \cap K \neq \emptyset.$$
\end{enumerate}
\end{theorem}

\emph{Remark 1:} Part (ii) asserts that away from the edges of $[-3,3]$ the consecutive spacing between elements of $K$ is at most $2$. One can prove the latter without the restriction on the location, but since we will not use this and the proof is much more cumbersome, we do not give it. See the end of Section \ref{sec:maxgapproofs} for a discussion.

Armed with these, one can formulate optimization/variational problems seeking maximal gap intervals, and more generally gap sets. Exact solutions to such optimization problems appear to be rare, one being the celebrated Alon-Boppana bound \cite{Ni} which, when combined with the existence of cubic Ramanujan graphs \cite{Ch}, is equivalent to $(2 \sqrt{2}, 3)$ being a maximal gap interval. Another maximal gap interval is $[-3,-2)$ as shown recently in \cite{K-F-S-H} using the results in \cite{C-G-S-S}. To these we add:

\begin{theorem}[    ]
\label{thm:extremalgaps}
$(-1,1)$  and $(-2,0)$ are maximal gap intervals. The first can be achieved with planar graphs and the second with planar multigraphs.
\end{theorem}

\emph{Remark 2:} By a multigraph we mean a graph with possible multiple edges between vertices or loops at vertices. That $(-1,1)$ is a maximal gap interval when restricted to bipartite cubic graphs was established in \cite{MoharMEdian, G-M}. Their examples achieving this gap are the same as our ``Hamburger'' graphs $W_b(n)$, which are bipartite and non-planar. These examples are shown in Fig. \ref{fig:Hamburger} and will be discussed in detail in Section \ref{sec:examples}.  In addition to these, we construct planar graphs that achieve the gap $(-1,1)$. These graphs have four faces which are triangles, and the rest of the faces are hexagons. (See Fig. \ref{fig:capping}.)

The large graphs which are free of eigenvalues in the intervals in Theorem \ref{thm:extremalgaps} are constructed as quotients of infinite cyclic covers of judiciously chosen base graphs. (See Section \ref{sec:examples} and Fig. \ref{fig:extremals}.)
According to Theorem \ref{thm:stronggap}(\textit{ii}) these intervals are maximal, which is proved in Section \ref{sec:maxgapproofs} using combinatorial constructions of approximate eigenfunctions.


The proof of Theorem \ref{thm:planargaps}, as well as our pursuit of further explicit extremal gap sets, makes use of the triangle map $\TT$ from $\XX$ to $\XX$, introduced in Ref. \cite{K-F-S-H} and which is investigated further in Section \ref{sec:T}. Given $Y \in \XX$, $\TT(Y)$  is a cubic graph obtained from $Y$ by replacing each vertex of $Y$ with a triangle and joining correspondingly. 
$\sigma(Y)$ and $\sigma(\TT(Y))$ are related by a simple formula involving the quadratic map $f(x) = x^2 -x -3$ (see Section \ref{sec:T}). The spectra of iterates of $\TT$ are captured by the dynamical properties of the iterates of $f$. Let
\begin{equation}\label{eqn:cantorset}
\Lambda = \bigcap_{m = 0} ^{\infty} {f^{-m} \left( [-3,3]   \right)}.
\end{equation}
Then $\Lambda$ is a Cantor subset of $[-3,3]$ which is $f$-invariant and on which the restriction of $f$ is a shift on two symbols (see Section \ref{sec:T}). The set
\begin{equation}\label{eqn:fullspec}
A := \Lambda \bigcup {\left(     \bigcup_{m = 0} ^{\infty} {f^{-m} \left( 0  \right)}   \right)}
\end{equation}
is a \textit{closed} subset of $[-3,3]$ consisting of the Cantor set $\Lambda$ and the isolated points $  \cup_{m = 0} ^{\infty} {f^{-m} \left( 0  \right)}$.

\begin{theorem}[ ]
\label{thm:orbits}
The set $A$ is a minimal planar spectral set and all the $Y$'s in $\XX$ for which $\sigma(Y) \subset A$ lie within finitely many $\TT$-orbits, moreover, $\C(A) = 1 $.

\end{theorem}

The triangle adding map $\TT$ allows us to construct new minimal spectral sets from old ones since $f^{-1}(K) \cup \{ 0,-2 \}$ is a such a set if $K$ is (see Proposition \ref{prop:Kspectral} in Section \ref{sec:T}), and the following tables record the basic extremal spectral/gap sets that we know of.

\begin{table}[ht]
\caption[A]{}
\centering
\textbf{Maximal Gap Intervals $\II$}

\vskip 0.1in
\begin{tabular}{ >{\centering\arraybackslash}m{1.7cm}||  >{\centering\arraybackslash}m{2.0cm}| >{\centering\arraybackslash} m{2.5cm}| >{\centering\arraybackslash}m{1.9cm}| >{\centering\arraybackslash}m{2.2cm}}
$\ \ \ \mathcal{I}:\ \ \ $ &  $\ \ \ [-3,-2)\ \ \ $ & $\ \ \  (-2,0)\ \ \ $ & $\ \ \ (-1,1)\ \ \ $   &  $\ \ \ (2\sqrt{2}, 3)$  \\ 
  \hline
  Properties:      &      Can be achieved by planar $Y$'s.    &    Can be achieved by planar multigraph $Y$'s      &    Can be achieved by planar $Y$'s     &    Cannot be achieved by planar $Y$'s. \\
\end{tabular}
\label{table:gapints}
\end{table}



\begin{table}[ht]
\caption[A]{}
\centering
\textbf{Minimal Spectral Sets $K$}

\vskip 0.1in

\begin{tabular}{ >{\centering\arraybackslash}m{2cm}|| >{\centering\arraybackslash}m{3.2cm}| >{\centering\arraybackslash}m{2.5cm}}
$\ \ \ K:\ \ \ $     & $\ \ \ [-2\sqrt{2},2\sqrt{2}] \cup \{3\}\ \ \ $       & $\ \ A$ \\ 
  \hline
Properties:      &  Cannot be achieved with planar $Y$'s.     &     Is achieved with planar $Y$'s.\\ 
\end{tabular}
\label{table:specsets}
\end{table}

Theorems \ref{thm:stronggap} and \ref{thm:orbits} show that $A$ has minimal capacity among all the minimal spectral sets, and we conjecture that the other entry in Table \ref{table:specsets} has maximal capacity.

Our results show that there are restriction on the size and structure of spectral sets of cubic graphs, but at the same time these sets are rich and complicated. It is interesting to compare this to similar questions that have been examined in other contexts. The eigenvalues of the Frobenius endomorphism on an Abelian variety over a finite field $\mathbb{F}_q$ are known to lie on the circle $C_q = \{ z \in \mathbb{C} ; |z| = \sqrt{q} \}$. In \cite{Se1} Serre extends the converse to Fekete's theorem \cite{Fe} and shows that in order for a closed conjugation-invariant subset $E$ of $C_q$ to contain the eigenvalues of a growing sequence of such Abelian varieties, it is essentially necessary and sufficient that the capacity of $E$ is at least $q^{1/4}$. (Note that $\C(C_q) = q^{1/2}$.) Thus, in this setting the capacity bound is the only restriction on the analog of being spectral. On the other hand, if one restricts to Abelian varieties that are Jacobians of curves over $\mathbb{F}_q$, then things rigidify and no gaps can be created, that is the minimal spectral set is $C_q$ itself \cite{T-V}.

Another setting in which the analog of the spectral set problem has been studied is that of locally symmetric spaces of rank bigger than one \cite{Seven}. The main result of Ref. \cite{Seven} implies that a sequence of such compact manifolds whose volume goes to infinity must Benyamini-Schramm converge to the universal cover. 
This is turn implies that their spectra (for the Laplacian and the full ring of invariant differential operators) become dense in the support of the Plancharel measure. In particular, there are no gap sets, and these spaces are spectrally (as well as in many other senses) very rigid. If the rank of the locally symmetric space is one, for example the case of compact hyperbolic surfaces, Selberg's eigenvalue conjecture \cite{Sar} implies that $(0, 1/4)$ is a gap set for the Laplacian, i.e. that there is a sequence of such surfaces whose areas go to infinity and which are free of Laplacian eigenvalues in $(0, 1/4)$. 
Interestingly, the question of whether all, or any, points in $[1/4, \infty)$ are gapped does not appear to have been addressed. This case is closest to our cubic graphs and at least constructions with Abelian covers might yield some examples.

Finally, while we have shown that spectral sets for planar cubic graphs are rich, these can become rigid if certain restrictions are imposed. For example, if a forthcoming paper \cite{K-W-S} with Fan Wei we show that for planar cubic graphs which have at most $6$ sides per face, $[-3,-1]\cup [1,3]$ is the unique minimal spectral set.

We end the introduction with a brief outline of the paper. In Section \ref{sec:T} we analyze the triangle map $\TT$ and its dynamics, as well as that of $f$ on the corresponding spectra. We prove Theorem \ref{thm:orbits}, and also Theorem \ref{thm:planargaps} assuming the results established in Section \ref{sec:examples}. In Section \ref{sec:coversandwaves} we review the theory of covering spaces and in particular the character torus (Brillouin zone) which parametrizes Abelian covers on which Bloch-wave spectral analysis takes place. Section \ref{sec:examples} is at the center of the paper, giving constructions of gap intervals. Applying the theory in Section \ref{sec:coversandwaves} to special $\ZZ$ and $\ZZ\times \ZZ$ covers of certain base $Y$'s that were found by numerical search, yields an arsenal of cyclic covers with exotic and even extremal gap intervals. In Section \ref{sec:maxgapproofs} we establish Theorems \ref{thm:stronggap} and \ref{thm:extremalgaps}. In Section \ref{sec:conclusion} we elaborate on the entries in Tables \ref{table:gapints} and \ref{table:specsets} and examine further extremal gap sets obtained using $\TT$.

\section{The Map $\mathcal{T}$}\label{sec:T}

\subsection{Definition and Properties of $\mathcal{T}$}\label{subsec:Tdef}

\begin{figure}[h]
	\begin{center}
		\includegraphics[width=1.0\textwidth]{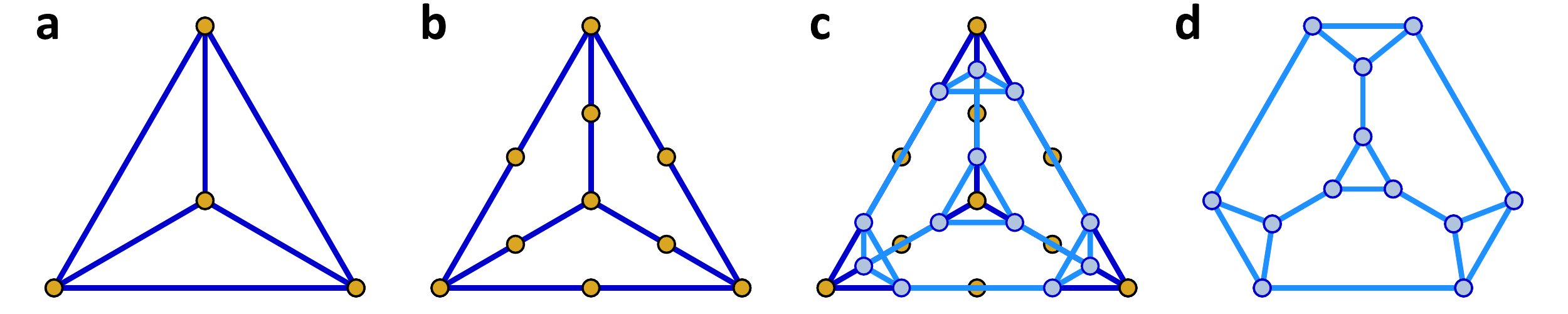}
	\end{center}
	\vspace{-0.3cm}
	\caption{\label{fig:Y4} 
     \textbf{Action of the map $\TT$.} Illustration of the action of the map $\TT$ using the complete graph on $4$ vertices $Y_4$, shown in \textbf{a}. \textbf{b} The subdivision $\mathcal{S}(Y_4)$. \textbf{c} The line graph $\TT(Y_4) = L(\mathcal{S}(Y_4))$ overlaid on subdivision graph.  Subdivision graph shown in dark blue, with gold vertices. Line graph indicated in light blue. This realization has crossing edges because the line-graph vertices are drawn at the midpoints of the original edges. \textbf{d} Alternate realization showing that $\TT(Y_4)$ is planar.
    } 
\end{figure}

Given $Y \in \XX$ let $\TT(Y)$ be the graph obtained as the composite of two operations: subdivide $Y$ into $\mathcal{S}(Y)$ by adding new vertices at the midpoints of the edges of $Y$, and then form the line graph of the $3,2$-biregular graph $\mathcal{S}(Y)$ to obtain $\TT(Y)$. This progression is illustrated in Fig. \ref{fig:Y4}. Put another way, $\TT$ replaces every vertex of $Y$ with a triangle and joins the corresponding edges between the triangles. Some immediate properties of $\TT$ are:
\begin{enumerate}\label{eqn:Tpropsi}

\item[(i)]   $ \TT : \XX \rightarrow \XX$

\item[(ii)]   $ |\TT (Y)|  = 3 |Y|$ (here  $|G|$ is the number of vertices of G)

\item[(iii)] If  $Y $ is planar, then so is $\TT(Y)$.

\end{enumerate}


For our purposes, the important properties of $\TT$ concern the relation of $\sigma(Y)$ to $\sigma(\TT(Y))$. 
\begin{enumerate}
\item[(iv)] The spectrum of $\TT(Y)$ is related to the spectrum of $Y$ by
\begin{equation}\label{eqn:Tpropsiv}
\sigma(\TT(Y)) = f^{-1} \left( \sigma(Y)   \right)\  \bigcup \ \{ 0 \} ^{n/2} \ \bigcup\  \{ -2 \} ^{n/2},
\end{equation}

\item[(v)] There exists $C_0 < \infty$ (one can give it explicitly) such that for $|Y| \geq C_0$
\begin{equation}\label{eqn:Tpropsv}
 Y = \TT(Z) \mbox{ for some } Z \iff \sigma(Y) \subset [-2,3]. 
\end{equation}

\end{enumerate}
The last gives a spectral characterization of the image of $\TT$; note that according to \ref{eqn:Tpropsiv},  $-2 \in \sigma(Y)$ for such $Y$'s. 
Relation (iv) is well known  (See \cite{H-S}), while (v) was derived an exploited in our recent paper \cite{K-F-S-H} and make use of the characterization of graphs whose spectra are contained in $[-2,\infty)$ (``Hoffman" graphs) \cite{C-G-S-S}.

\begin{figure}[h]
	\begin{center}
		\includegraphics[width=0.8\textwidth]{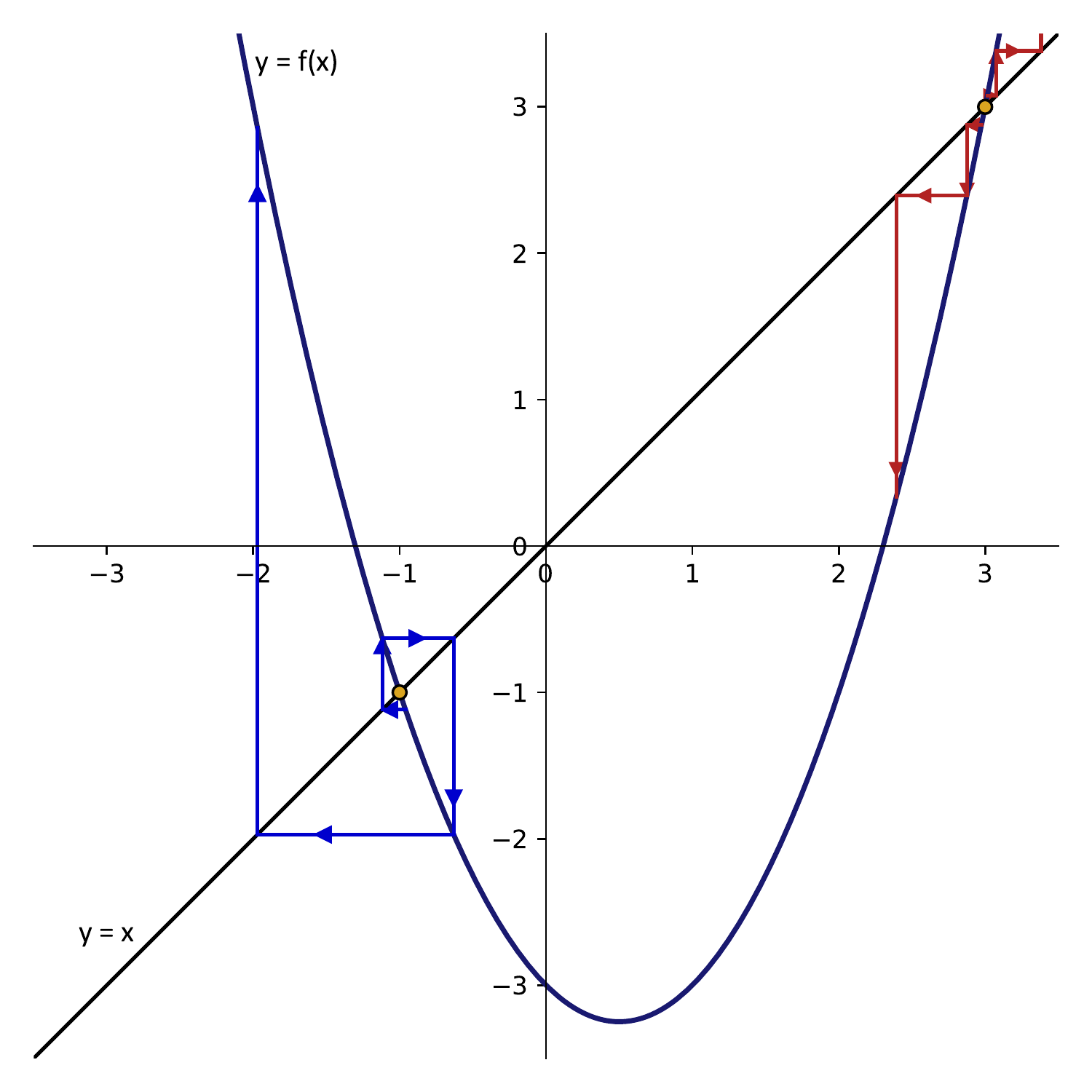}
	\end{center}
	\vspace{-0.6cm}
	\caption{\label{fig:flowFig} 
     \textbf{Fixed points of $f$.} Plot of $f(x)$ showing the flow that occurs under iteration of $f$. The two fixed points $-1$ and $3$ are indicated by gold dots. Both of these fixed points are unstable. Sample flows are shown in blue and red, respectively. For an initial $x_0$ near the fixed point, its image $ x_1 = f(x_0)$ is farther away, and iteration rapidly runs off to infinity. } 
\end{figure}

$\TT$ provides us with a versatile tool to construct spectral sets.

\begin{proposition}\label{prop:Kspectral}
\begin{enumerate}
\item[] 

\item[(A)] If $K$ is a spectral set, then so is $f^{-1} (K) \cup \{0,-2 \}$.

\item[(B)] If $K$ is a minimal spectral set, then so is $f^{-1} (K) \cup \{0,-2 \}$.

\end{enumerate}
\end{proposition}
\textit{Proof:} (A) $K$ is spectral means that there is a sequence of $Y_n \in \XX$ with $|Y_n| \rightarrow \infty$ such that $\sigma(Y_n) \subset K$. From \ref{eqn:Tpropsiv} it follows that $\sigma(\TT(Y))$ is a contained in $f^{-1} (K) \cup \{0,-2 \}$, so that the latter is spectral.

(B) To show that $f^{-1} (K) \cup \{0,-2 \}$ is minimal, it suffices to show that if $Y_n \in \XX$ with $|Y_n| \rightarrow \infty$ has $\sigma(Y_n) \subset f^{-1}(K) \cup \{ -2,0\}$, then 
$$ \overline{\bigcup_{n}{\sigma(Y_n)} } = f^{-1}(K)\  \bigcup \ \{-2,0\}.$$
Now $f^{-1}[-3,3] = [-2,0] \cup [1,3]$, and $Y_n$ is therefore a Hoffman graph. Hence, according to \ref{eqn:Tpropsv}, $Y_n = \TT(Z_n)$ for $n$ large enough. Moreover, $f^{-1}(\sigma(Z_n)) \subset f^{-1}(K)$, and hence, $\sigma(Z_n) \subset K$. Since $K$ is minimal, it follows that 
$$ \overline{\bigcup_{n}{\sigma(Z_n)} } =K.$$
Thus, 
$$ f^{-1} \left(\overline{\bigcup_{n =1}^\infty{\sigma(Z_n)} } \right) = f^{-1}(K).$$
Now
$$ f^{-1} \left(\overline{\bigcup_{n =1}^\infty{\sigma(Z_n)} } \right) = \overline{ f^{-1} \left(\bigcup_{n =1}^\infty{\sigma(Z_n)}  \right) } ,$$
and 
$$ f^{-1} \left( \bigcup_{n =1}^\infty{\sigma(Z_n)}  \right)  = \bigcup_{n = 1}^\infty{f^{-1} \left( \sigma(Z_n)  \right)}.$$
Thus, 
$$  \overline{ f^{-1} \left(\bigcup_{n =1}^\infty{\sigma(Z_n)}  \right) } \bigcup \ \{ -2,0 \} = f^{-1} (K) \ \bigcup\ \{ -2,0 \},$$
and 
$$ \overline{   \bigcup_{n}{\sigma(Y_n)} } = f^{-1}(K) \ \bigcup\ \{-2,0\},$$
as required.

\begin{figure}[h]
	\begin{center}
		\includegraphics[width=1.0\textwidth]{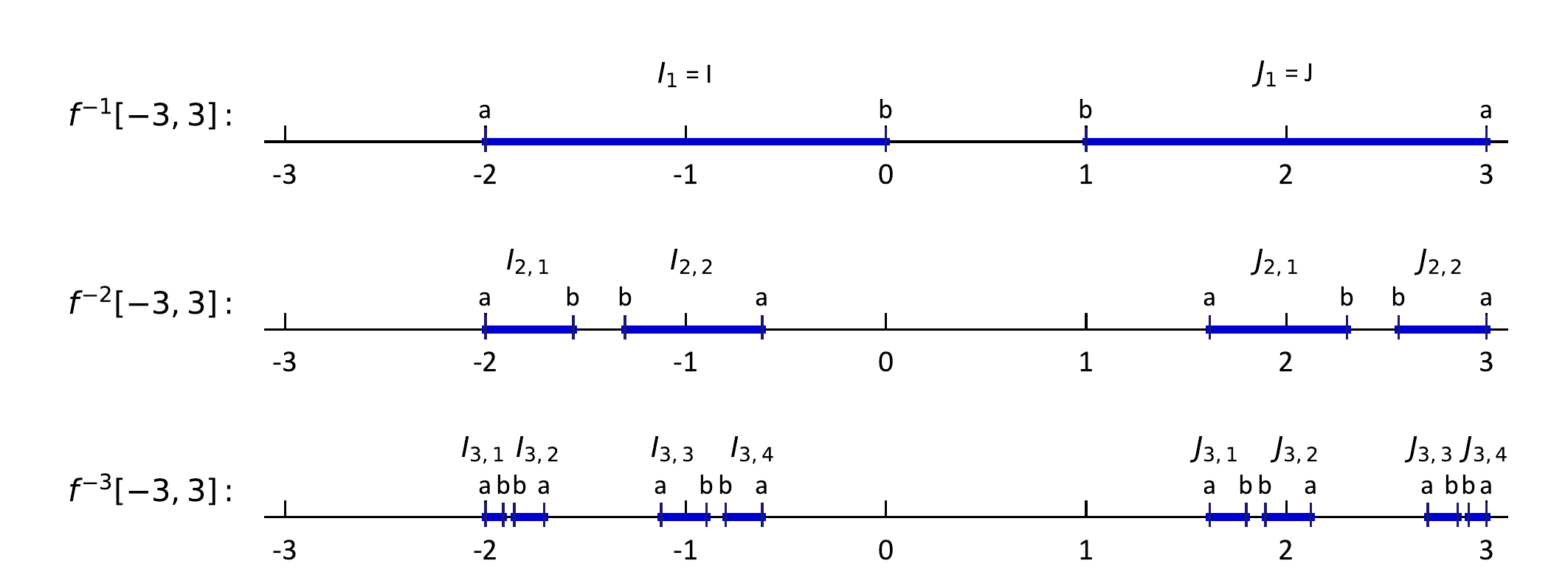}
	\end{center}
	\vspace{-0.3cm}
	\caption{\label{fig:IntervalFlowFig} 
     \textbf{Action of $f^{-1}$.} As described in the main text, the function $f^{-1}$ acting on $[-3,3]$ divides it into two image intervals $I$ and $J$. Successive application of $f^{-1}$ causes further fragmentation. The images $f^{-m}([-3,3])$ are shown above for $m = 1,2,3$. The image intervals are indicated as thick, dark blue, lines, and labeled $I_{m,j}$ and $J_{m,j}$. Each such interval has two end points, labeled by $a$ and $b$. The $a$-type endpoints are $f^{-m}(3)$ and once they appear remain fixed under further application of $f^{-1}$. The $b$-type endpoints are $f^{-m}(-3)$ and are successively removed each time. Therefore, the $a$'s are in $\Lambda$, but the $b$'s are not.
    } 
\end{figure}

To go further, we examine in more detail the dynamical properties of iterating $f$ as a map from $\mathbb{R}$ to $\mathbb{R}$. The fixed points $x = f(x)$ of $f$ are $x = -1$ and $x = 3$, and both are repelling, as illustrated in Fig. \ref{fig:flowFig}. The forward orbit of any $x \in \RR$ is $\infty$ except for a closed Cantor subset $\Lambda$ of $[-3,3]$ that we describe below (see \cite{De} for a detailed discussion).
Note that $I = [-2,0]$ and $J = [1,3]$ are mapped homeomorphically to $[-3,3]$ under $f$. Here the external end points $\{-2,3\}$ correspond to $f^{-1}(3)$ and the internal ones $\{0,1\}$ to $f^{-1}(-3)$. All points outside of $I$ or $J$ are mapped after iteration to $(3,\infty)$ and hence after further iterations to $\infty$. Repeating this cuts out two subintervals in each of $I$ and $J$, outside of which $f^2$ maps to $(3,\infty)$.

$f^{-m}([-3,3])$ consists of $2^m$ intervals, half of them to the left of $0$ and half to the right of $1$, symmetrically placed about $1/2$. 
As sketch of these intervals for $m = 1,2,3$ is shown in Fig. \ref{fig:IntervalFlowFig}.
The intervals $I_{m,j} , \  j = 1,2,\cdots, 2^{m-1}$ and $J_{m,j}, \ j = 1,2,, \cdots 2^{m-1}$ have end points denoted by $a \in f^{-m}(3)$ and $b\in f^{-m}(-3)$. Once an $a$ endpoint appears for some $m$, it remains an endpoint for all levels $t \geq m$. It follows that the points $f^{-m}(3)$ all lie in $\Lambda$ where 
\begin{equation}\label{eqn:Lambdadef}
\Lambda = \bigcap_{m=1}^{\infty} {f^{-m} \left( [-3,3] \right) }.
\end{equation}
Note that $f^{-1}\left( [-3,3]\right) \supset f^{-2}\left( [-3,3]\right) \supset f^{-3}\left( [-3,3]\right) \supset \cdots$, and that $\Lambda$ is a non-empty Cantor subset of $I\cup J$. For $\xi \in \Lambda$, define $s (\xi)$ in the compact topological space $F = \{0,1\}^\mathbb{N}$ by $s(\xi) = (x_1, x_2, x_3, \cdots )$, where $x_j = 0$ if $f^{j}(\xi) \in I$ and $x_j = 1$ if $f^{j}(\xi) \in J$. Then $s : \Lambda \rightarrow F$ is a homeomorphism and it conjugates the action of $f|_\Lambda$ to the shift $\bf{S}$ given by $\bf{S} \rm : (x_1, x_2, x_3, \cdots) \mapsto (x_2, x_3, \cdots )$ (see \cite{De}). That is
\Large
\begin{equation}\label{eqn:commutingflow}
\begin{gathered}
\xymatrixcolsep{3pc}
\xymatrixrowsep{2pc}
\xymatrix{
\Lambda \ar[d]^s \ar[r]^f & \Lambda \ar[d]^s\\
F \ar[r]^{\bf{S}\rm}          &F}
\end{gathered}
\end{equation}
\normalsize
is a commuting diagram.

The dynamics of $f: \RR \rightarrow \RR$ consists of two very different behaviors. On $\RR \backslash \Lambda$ every forward orbit of a point goes to $\infty$, while on the invariant set $\Lambda$ the action of $f$ is chaotic, being conjugate to a one-sided shift on the two-point infinite sequence space $F$. Note that the two fixed points of $f$, $-1$ and $3$, correspond to the fixed points $(0,0,0, \cdots)$ and $(1,1,1,\cdots)$ of $\bf{S}$, respectively.

To make use of \ref{eqn:Tpropsiv}  when iterating $\TT$, we need to keep track of $f^{-m}(-2)$ and $f^{-m}(0)$ for $m = 1, 2, 3, \cdots$. Now $-2 \in \Lambda$ as is $\cup_{m=0}^\infty{f^{-m}(-2)}$, these being the ``$a$-endpoints'' of the intervals defining $\Lambda$. On the other hand $0 \notin \Lambda$ and its pre-periodic points $\cup_{m=0}^\infty{f^{-m}(0)}$ consist of half of the ``$b$-endpoints" of the intervals defining $f^{-m}([-3,3])$.
While none of these points are in $\Lambda$, the limit points of $\cup_{m=0}^\infty{f^{-m}(0)}$ are all contained in $\Lambda$, 
as is clear from its definition. Hence if 
\begin{equation}\label{eqn:Adef}
A := \Lambda \ \bigcup\ \left( \bigcup_{m = 0}^\infty{ f^{-m}(0)} \right), 
\end{equation}
then $A$ is a closed subset of $[-3,3]$, which consists of the Cantor set $\Lambda$ and the infinite set $\cup_{m=0}^\infty{f^{-m}(0)}$ of isolated points.

\setcounter{manualtheorem}{3}
\begin{manualtheorem}
$A$ is a minimal planar spectral set, and all $Y$'s in $\XX$ for which $\sigma(Y) \subset A$ lie in finitely-many $\TT$-orbits, moreover, $\C(A) = 1$.
\end{manualtheorem}
\textit{Proof:} Let $Y_4$ be the $3$-regular graph on four vertices shown in Fig.~\ref{fig:Y4} \textbf{a}. $\sigma(Y_4) = \{-1\}^3 \cup \{3 \}$, and since both $-1$ and $3$ are in $\Lambda$, it follows from \ref{eqn:Tpropsiv} and the $f$-invariance properties of $\Lambda$ and of $A$ that $\sigma(\TT^m(Y_4)) \subset A$ for $m \geq 0$. This shows that $A$ is spectral, and planar since $\TT^m(Y_4)$ is. To see that it is minimal, let $Y_n \in \XX$ with $|Y_n| \rightarrow \infty$ and $\sigma(Y_n) \subset B$ with $B\subset A$, a closed set. Since $\sigma(Y_n) \subset [-2,3]$, if follows from \ref{eqn:Tpropsv} that for $n$ large, $Y_n = \TT(Z_n)$ and $\sigma(Y_n) = f^{-1}(\sigma(Z_n)) \cup \{-2,0\}$. Now since $\sigma(Y_n)\subset B\subset A$, we have that $\sigma(Y_n)  \subset f^{-1}([-2,0]) \cup f^{-1}([1,3])$, so
$$ f^{-1}(\sigma(Z_n)) \subset  \Big[ f^{-1}([-2,0]) \cup f^{-1}([1,3]) \Big] .$$
It follows that 
$$ \sigma(Z_n) \subset [-2,3].$$
Hence, using \ref{eqn:Tpropsv} again we have that for $n$ large enough $Z_n = \TT(Z_{n'})$, that is $Y_n = T^2(Z_{n'})$ for $n$ large.

Repeating this argument $k$ times, we see that for $n$ large, $Y_n = \TT^k(Z_n)$. However, this then implies that 
$$ \sigma(Y_n) \supset  f^{-k}(-2) \ \bigcup\ f^{-k}(0) \ \bigcup \ \{3\}  .$$
Since this holds for all $k$, we have that 
\begin{equation}\label{eqn:containment}
A \supset B \supset \bigcup_{k = 0}^\infty { \left[   f^{-k}(-2) \ \bigcup\ f^{-k}(0)   \right] } \ \bigcup\ \{3 \}.
\end{equation}
From the description of the $f$-action on $\Lambda$ in terms of $\bf{S}$ on $F$, it is clear that $\cup_{k = 0}^\infty f^{-k}(-2)$ is dense in $\Lambda$, while $\cup_{k = 0}^\infty f^{-k}(0)$ covers the discrete part of $A$. Since $B$ is closed it then follows from Eqn. \ref{eqn:containment} that $B = A$. This proves that $A$ is a minimal spectral  set.

In the above argument, given $Y\in \XX$ with $\sigma(Y) \subset A$, we repeatedly found $Y = \TT(Z_1), Z_1 = \TT(Z_2), \cdots, Z_{j-1} = \TT(Z_j)$, as long as $|Z_{j-1}| > C_0$, where $C_0$ is the constant in \ref{eqn:Tpropsv}. It follows that $Y = \TT^k(Z)$ for  $Z$ in the finite set of graphs in $\XX$ with $|Z| \leq C_0$ and some $k \geq 0$. This proves the second part of Theorem \ref{thm:orbits}, namely that any such $Y$ lies in a finite number of $\TT$-orbits. The number of such orbits is the number of $\TT$-inequivalent $Y$'s with $|Y| \leq C_0$ and $\sigma(Y) \subset A$. One such orbit is that of $Y_4$ shown above. In addition to this, we know of two more which are given in Section \ref{subsec:multigraphs}.

To complete the proof of the last statement in Theorem \ref{thm:orbits} we need to compute the capacity of $A$. Since the points of $A$ which are not in $\Lambda$ are isolated, it follows that $\C(A) = \C(\Lambda)$. We apply (5.2) of Theorem 11 of \cite{G-V} which, when applied to our $f$, yields that if 
$$ E_0 \subset [-3,3] \mbox{ is closed}$$
and $E_1 = f^{-1}(E_0)$ then $\C(E_1) = \sqrt{\C(E_0)}.$
Applying this to $E_0 = [-3,3]$  we have that $\C(E_0) = 3/2$, (the capacity of an interval of length $L$ is $L/4$) and hence 
$$ \C \left(  f^{-1} ([-3,3])  \right) = \sqrt{\frac{3}{2}} .$$
Applying this repeatedly yields that 
$$\C \left(  f^{-m} ([-3,3])  \right) = \left(   \frac{3}{2} \right) ^{1/2^m} . $$
Letting $m \rightarrow \infty$ we get that 
$$ \C \left( \bigcap_m  f^{-m} ([-3,3])  \right) = \C(\Lambda) = 1.$$


Before moving on to consider multigraphs which lead to additional $\TT$-orbits, we apply the triangle map $\TT$ to show that every $\xi \in [-3,3)$ is planar gapped, that is that for every such $\xi$ there is a neighborhood $U$ of $\xi$ which is a planar gapped set.
\setcounter{manualtheorem}{0}
\begin{manualtheorem}\label{} 
Every $\xi$ in $[-3,3)$ is planar gapped.
\end{manualtheorem}
\textit{Proof:} In Section \ref{subsec:planargaps} we present some special Abelian covers which show by explicit constructions that every point $\xi \in [-3, 2\sqrt{2}]$, and in particular, every point of $I$, is planar gapped. We use $\TT$ to deal with the remaining points:

Note first that if $f(\xi)$ is planar gapped, then so is $\xi$. Indeed, according to the above $-2$ and $0$ are in $I = [-2,0]$ and are planar gapped. Let $U$ be a neighborhood of $f(\xi)$ which is gapped, witnessed by a sequence $Y_m$ which $\sigma(Y_m) \cap U = \emptyset$. Since $\sigma(\TT(Y_n)) = f^{-1} (\sigma(Y_n)) \cup \{0,-2\}$, 
we have that 
$$ f^{-1} \left( \sigma(Y_n) \right)  \ \bigcap \ f^{-1}(U) = \emptyset,$$
and $\xi \in f^{-1}(U)$.

Let $V$ be a neighborhood of $\xi$ contained in $f^{-1}(U)$ and not containing $0$ or $-2$ (which we can assume since by the remark above $\xi \notin \{0,-2\}$ as these two points are gapped), then 
$$ \xi \in V \mbox{ and } V\ \bigcap \ \left(  f^{-1}\left( \sigma(Y_n) \right) \ \bigcup\ \{-2,0\}       \right) = \emptyset.$$
Hence, $\TT(Y_n)$ certifies that $\xi$ is gapped, and in fact planar gapped. Iterating this argument $k$ times yields:
\begin{equation}\label{eqn:planargapstatement}
\mbox{ If } f^k(\xi) \mbox{ is planar gapped, then so is } \xi.
\end{equation}
Applying \ref{eqn:planargapstatement} to the point $0$, which is already known to be planar gapped, we conclude that:
\begin{equation}\label{eqn:planargapstatement2}
\mbox{ Any point in }  f^{-m}(0), m \geq 0 \mbox{ is planar gapped}.
\end{equation}

Any point not in $A$ is planar gapped as witnessed by $\TT^m(Y_4), m \geq 0 $. \ref{eqn:planargapstatement}  and \ref{eqn:planargapstatement2} leave the points in $\Lambda$ as the only ones which have not as yet been shown to be gapped. Now, if $\xi \in \Lambda$ and $\xi \neq 3$, then $s(\xi) = (x_1,x_2, x_3, \cdots)$ with at least one $j \geq1$ having $x_j = 0$, then $f^j(\xi) \in I$ for that $j$. Now all points of $I$ are planar gapped, so by applying  \ref{eqn:planargapstatement} it follows that $\xi$ is gapped. This completes the proof of Theorem \ref{thm:planargaps}.

\begin{figure}[h]
	\begin{center}
		\includegraphics[width=1.0\textwidth]{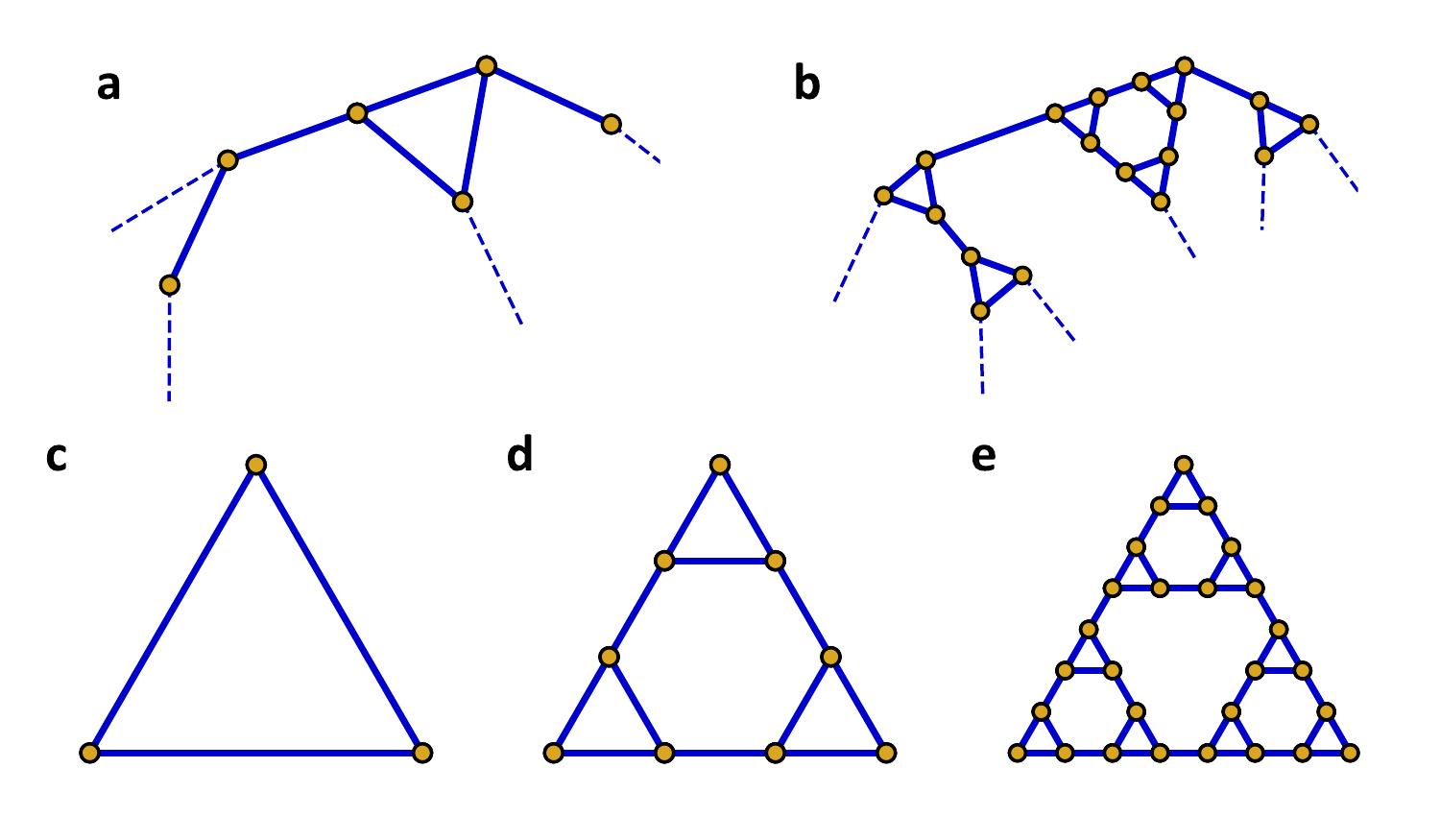}
	\end{center}
	\vspace{-0.6cm}
	\caption{\label{fig:Hanoi} 
     \textbf{Iterates of $\TT$ and Hanoi graphs.} \textbf{a} A typical neighborhood for a graph $Y \in \XX$. Dashed edges show some of the edges leaving the neighborhood. Under application of $\TT$ vertices are converted into triangles, and a typical neighborhood resembles \textbf{b}.  Further application of $\TT$ transforms vertices into the Hanoi graphs for 3 pegs $H_3^m$. \textbf{c}-\textbf{e} show these graphs for $m = 0,1,2$, respectively.
    } 
\end{figure}


We end this section with some remarks about the shapes of large iterates of $\TT$. Given an initial $Y\in \XX$, applying $\TT$ $m$ times replaces the vertices of each successive generation with triangles. Sketches of characteristic neighborhoods in $Y$ and $\TT^m(Y)$ are shown in Fig. \ref{fig:Hanoi} \textbf{a}, \textbf{b}. Alternatively, one can look directly at $\TT^m(Y)$ and look at the larger neighborhood that arises from each original vertex of $Y$. For one iteration, the new neighborhood is a simple triangle, more generally the structure that appears is a tower of Hanoi graph for $3$ pegs and $m$ discs $H_3^m$ \cite{G-S}, or equivalently, the infinite Sierpinski pre-lattice in \cite{H-S}.
The progression of these graphs is shown in Fig. \ref{fig:Hanoi} \textbf{c}-\textbf{e}.

Thus, for large $m$, the local shape of $\TT^m(Y)$ is dictated by that of $H_3^m$. The spectra of the graphs $H_3^m$ with loops added at the three vertices of the outer triangle were computed explicitly in \cite{G-S} and \cite{H-S}.
Not surprisingly, our minimal spectral set $A$ emerges as the closure of their spectra (see Theorem 1.1 \cite{G-S}). 

It would be interesting to extend the map $\TT$ to the closure of $\XX$ in the space of Benyamini-Schramm limits and graphings \cite{Lo} and to study the dynamics of $\TT$ on these spaces.

\subsection{Relevant Graphs with Multiple Edges and Loops}\label{subsec:multigraphs}

\begin{figure}[h]
	\begin{center}
		\includegraphics[width=1.0\textwidth]{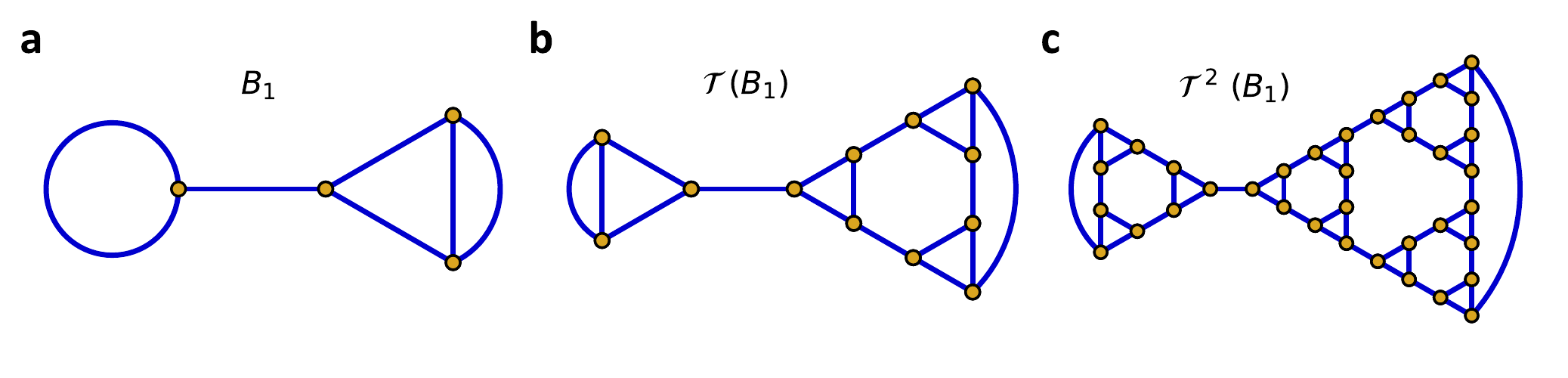}
	\end{center}
	\vspace{-0.6cm}
	\caption{\label{fig:Bbar} 
    \textbf{The multigraph $B_1$.} The multigraph $B_1$ and iterates of it under $\TT$.  \textbf{a} $\sigma(B_1)  = \{3,2,-1,-2\} \subset A$. \textbf{b} $\TT(B_1)$ still has multiple edges; however, $\TT^2(B_1)$, shown in \textbf{c}, does not. If $Y_1 = \TT^2(B_1)$, then $Y_1 \in \XX$ and $\sigma(Y_1) \subset A$. $\TT^m(Y_1)$ is a new $\TT$-orbit distinct from the orbit of $Y_4$ discussed in the text.
      } 
\end{figure}

\begin{figure}[h]
	\begin{center}
		\includegraphics[width=1.0\textwidth]{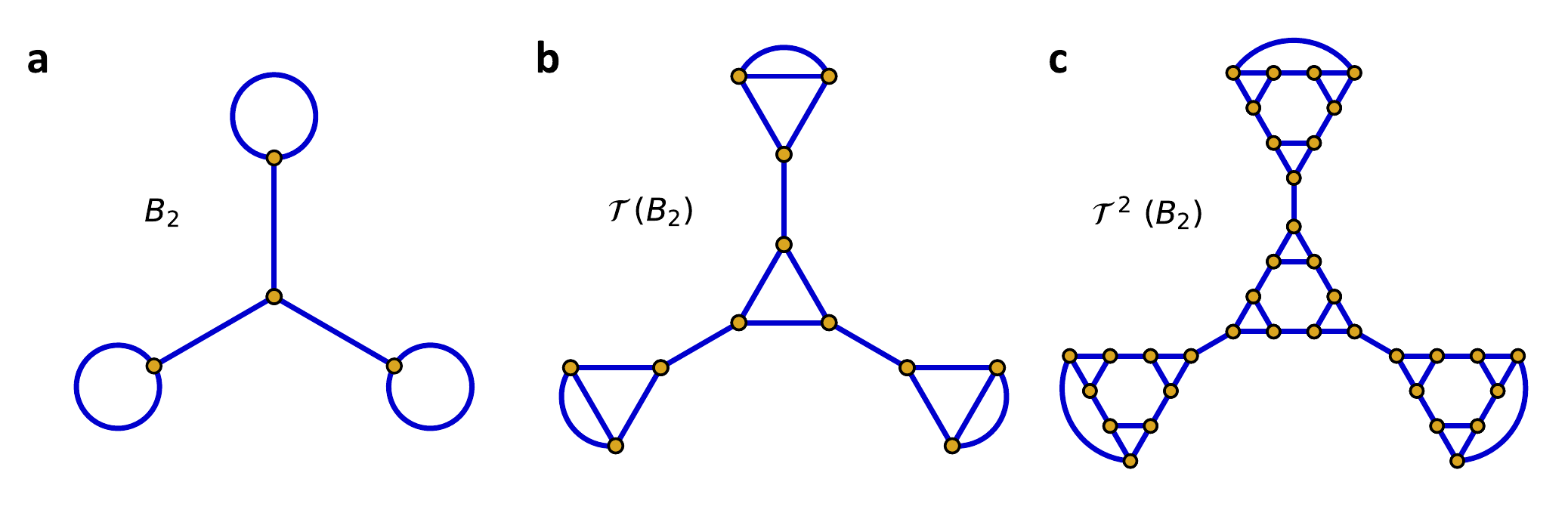}
	\end{center}
	\vspace{-0.6cm}
	\caption{\label{fig:Bstar} 
    \textbf{The multigraph $B_2$.} The multigraph $B_2$ and iterates of it under $\TT$.  \textbf{a} $\sigma(B_2)  = \{3,2,2,-1\} \subset A$. \textbf{b} $\TT(B_2)$ still has multiple edges; however, $\TT^2(B_2)$, shown in \textbf{c}, does not. If $Y_2 = \TT^2(B_2)$, then $Y_2 \in \XX$ and $\sigma(Y_2) \subset A$. $\TT^m(Y_2)$ is a new $\TT$-orbit distinct from the orbit of $Y_4$ discussed in the text, and that of $Y_1$ (shown in Fig.~\ref{fig:Bbar}).
    } 
\end{figure}

While we predominantly restrict to considering only graphs with single edges and no loops, there are several multigraphs which are extremely relevant. Two such examples are shown in Figs. \ref{fig:Bbar} - \ref{fig:Bstar}. The action of $\TT$ on such graphs is well defined, and after a few iterations of $\TT$, the resulting graph no longer possess any multiple edges or loops, as shown in Figs.~\ref{fig:Bbar} - \ref{fig:Bstar}. In this way, multigraphs can initiate orbits of $\TT$ in the space of graphs with no multiple edges or loops. Such is the case for $B_1$ and $B_2$ (Figs. \ref{fig:Bbar} - \ref{fig:Bstar}). The spectrum of both of these graphs is contained in $A$: $\sigma(B_1) = \{3,2,-1,-2\} \subset A$, and $\sigma(B_2) = \{3,2,2,-1\}$.
In fact, the images under $\TT$ of the multigraphs $B_1$ and $B_2$ constitute two new orbits, both distinct from that of $Y_4$ which was used above to establish the spectral properties $\TT$ and $A$ and prove Theorem \ref{thm:orbits}.
Another relevant multigraph will be discussed in Section \ref{subsec:extremalgaps} where it arises as the smallest possible primitive cell for an example Abelian cover which realizes the bipartite maximal gap interval $(-1,1)$.

\section{Abelian Covers and Bloch Waves}\label{sec:coversandwaves}

\subsection{General Covering Space and the Character Torus}\label{subsec:covers}

In order to construct small spectral sets, we examine large regular covers of a fixed graph. We review the general theory and then specialize so as to make explicit computations. For detailed exposition of the topological notions in the context of graphs see (\cite{Su} and \cite{S-T}).

If $Y$ is a finite connected graph, we can view it as a one-dimensional topological space. Let $\tilde{Y}$ be its universal cover and $\pi_1 = \pi_1(Y,y_0)$ the fundamental group of $Y$ based at $y_0$. $\pi_1$ is a free group on $k = n-m+1$ generators, where $m$ is the number of edges of $Y$ and $n$ the number of vertices (see below for explicit generators). 
Any $3$-regular graph can be constructed from the $3$-regular tree by equating vertices and ``stitching" branches together. The $3$-regular tree is thus the universal cover $\tilde{Y}$ of all $3$-regular graphs, and any such $Y$ by modding out the corresponding group of vertex automorphisms on the tree.
Thus $Y\cong \tilde{Y}/\pi_1$
 and finite covers $Z$ of $Y$ correspond to finite index subgroups $\Delta$ of $\pi_1$.
\begin{eqnarray}\label{eqn:maxcov}
 \tilde{Y} &  & \\ \nonumber
\downarrow &    &  \\ \nonumber
Z &=& \tilde{Y}/\Delta \\ \nonumber
\downarrow &    &  \\ \nonumber
Y & =   & \tilde{Y} / \pi_1  \\ \nonumber
\end{eqnarray}

If $\Delta$ is a normal subgroup of $\pi_1$, then $Z$ is a regular cover of $Y$ with deck (Galois) group $G = \pi_1/\Delta$ acting on the points of $Z$ lying over a given point in $Y$. Abelian covers of $Y$ are ones for which $G$ is Abelian and we also allow $G$ to be infinite. 
It is difficult to analyze the spectrum of a general regular cover $Z$ of $Y$, however for Abelian covers there is a torus that parametrizes such covers and allows for a finite analysis of their spectra. We restrict our attention to these.

The Abelian covers $W$ of $Y$ correspond to epimorphisms
\begin{equation}\label{eqn:epimorph}
\rho : \pi_1 \rightarrow G,
\end{equation}
where $G$ is an Abelian group generated by $k$-elements, and  $W = \tilde{Y} / (ker \rho)$. Any morphism $\rho$ in Eqn. \ref{eqn:epimorph} factors through the maximal Abelian quotient $H_1(Y, \ZZ)$;
\begin{equation}\label{homology1}
H_1(Y, \ZZ) = \pi_1/[\pi_1,\pi_1].
\end{equation}
$H_1(Y, \ZZ)$ is the first integral homology group and 
\large
\begin{equation} \label{eqn:mapping}
\begin{gathered}
\xymatrix{
\pi_1 \ar[r]^h \ar@/_1pc/[rr]_{\rho} & H_1 \ar[r]^{\beta} & G\\
}
\end{gathered}
\end{equation}
\normalsize
where $h$ is the Hurwitz quotient and $\beta$  a morphism of $H_1$ onto $G$.

If $W_{max} := \tilde{Y} / [\pi_1, \pi_1]$, then $W_{max}$ is an $H_1$ cover of $Y$ and it is the maximal Abelian such cover. If $W$ is an any Abelian cover of $Y$, then $W = W_{max}/B$ with $B$ a subgroup of $H_1$ and $W$ is an $H_1/B$ cover.
\begin{eqnarray}\label{eqn:maxabcov}
W_{max} & &  \\ \nonumber
\downarrow\ &&     \\ \nonumber
W \ & =& W_{max}/B \ , \ G(W/Y) = H_1/B. \\ \nonumber
\downarrow\ &&  \\ \nonumber
Y \ &=& W_{max} / H_1    \\ \nonumber
\end{eqnarray}
$H_1(Y, \ZZ)$ is a isomorphic to $\ZZ^k$ (see below for explicit generators) and the key to our analysis is its dual torus $T = T(Y)$
\begin{equation}\label{eqn:dualT}
T := (H_1(Y, \ZZ))^\wedge,
\end{equation}
that is the topological group of all characters $\chi : H_1 \rightarrow S^1$, $S^1  = \{ z \in \mathbb{C} : |z| = 1 \}$. $T$ is isomorphic to $(S^1)^k$. 
These entities are used extensively in crystallography solid state physics with different terminology. $T$ is equivalent to the Brillouin zone of $W_{max}$ and the characters correspond to the phases $\exp{i(\vec{k}\cdot \vec{x})}$ for quasimomenta $\vec{k}$, (see e.g. \cite{Gi}).

The spectra of the adjacency matrices of $W$ in Eqn \ref{eqn:maxabcov} can be analyzed through the spectra of the `$\chi$-twisted' operators:
\begin{equation}\label{eqn:Vchi}
V_\chi := \{ h : \ \ \ h: W \rightarrow \mathbb{C} ,\  h(\gamma x) = \chi(\gamma) h(x) \  \mbox{ for } \ x \in W_{max}, \gamma \in H_1 \}.
\end{equation}
$V_\chi$ is a linear space of dimension $n$ (the $h$'s determined by their values on the projection of $x$ to $Y$). The adjacency operator $\ad$ preserves $V_\chi$,
and we denote its restriction to $V_\chi$ by $\ad_\chi$. 
It can be checked that with respect to the standard inner product on $V_\chi$ (again see below in terms of a basis) that $\ad_\chi$ is hermitian and that its eigenvalues lie in $[-3,3]$. Let
\begin{equation}\label{eqn:eigslist}
-3 \leq \lambda_1 (\chi) \leq \lambda_2 (\chi) \cdots \leq \lambda_n (\chi) \leq 3
\end{equation}
denote the eigenvalues with multiplicities. As functions of $\chi$ we can choose the $\lambda_j$'s to be continuous on $T$.

If $B$ is a subgroup of $H_1$, the annihilator of $B$, denoted by $B^\perp$ is the closed subtorus
\begin{equation}\label{eqn:Bperp}
B^\perp = \{ \chi \in \hat{H}_1 : \chi(b) = 1 \mbox{ for all } b \in B \},
\end{equation}
and 
\begin{equation}\label{eqn:Bperpsentence}
 B^\perp \mbox{ is canonically isomorphic to } (H_1/B)^\wedge.
\end{equation}
The spectrum of any finite Abelian cover $W$ of $Y$ (as in Eqn. \ref{eqn:maxabcov}) is equal to 
\begin{equation}\label{eqn:bandunion1}
\sigma(W) = \bigcup_{j = 1}^{n} \lambda_j (B^\perp).
\end{equation}
It is convenient to extend this to any closed subgroup $S$ of $T$ (such a subgroup is a finite union of connected subtori and translates thereof by finitely many torsion points) setting
\begin{equation}\label{eqn:bandunion2}
\lambda(S) := \bigcup_{j = 1}^n \lambda_j(S).
\end{equation}
$\lambda(S)$ is a finite union of closed intervals in $[-3,3]$ called bands.

Our construction of spectral sets proceeds by choosing $S$ to be infinite but small, in fact,  one dimensional. It contains arbitrarily large finite subgroups which we can take to be $B^\perp$'s, and then, according to Eqn. \ref{eqn:bandunion1}, we have that for any such $W$ whose $B^\perp$ is contained in $S$,
\begin{equation}\label{eqn:pbcspectrum}
\sigma(W) \subset \lambda(S),
\end{equation}
and hence $\lambda(S)$ is a spectral set.
Note that since $\chi = 1$ is in $S$,

\begin{equation}\label{eqn:k0eigs}
\lambda(S) \supset \sigma(Y).
\end{equation}
Our engineering of extremal spectral sets, which we describe in detail in Section \ref{sec:examples}, is to start with a $Y$ with suitable gaps in its spectrum and then to search for special one-dimensional subtori $S$ of $T$ for which the inclusion in Eqn. \ref{eqn:k0eigs} is as tight as possible.
It cannot be too tight since according to Theorem \ref{thm:stronggap} the capacity of $\lambda(S)$ is at least $1$.

The extension in Eqn. \ref{eqn:bandunion2} can be interpreted in terms of the spectral theory of $\ad$ on infinite Abelian covers of $Y$, often referred to as Bloch wave theory in this setting. If $W$ in Eqn. \ref{eqn:maxabcov} is an $H_1/B$ Abelian cover of $Y$ and $\sigma(W)$ is the $\ell^2$-spectrum of $\ad$, where $\ad$ is a self-adjoint operator on the natural $\ell^2$-space of functions on $W$, then Bloch wave theory \cite{R-S, Gi} yields
\begin{equation}\label{eqn:Aspec}
\sigma(A_W) = \lambda(S), \mbox{ where } S = B^\perp,
\end{equation}
and Eqn. \ref{eqn:bandunion2} is the familiar band structure of the spectrum.

The torus $T$ also gives an alternative description of the Abelian covers $W$ in Eqn. \ref{eqn:maxabcov}. Each closed subgroup $S$ of $T$ determines a $W$ by taking $B = S^\perp = \{ b \in H_1 : \chi(b) = 1 \mbox{ for all } \chi \in S \}$ and $W = W_{max}/B$. $S$ is then canonically the dual group of this $H_1/B$ cover of $Y$.
 See Section \ref{sec:examples} for an explicit description of how to construct such constrained covers graphically.

\subsection{Explicit Bases and Flat Bands}\label{subsec:bases}
In order to make fruitful computations, we need to choose generators for $H_1(Y,\ZZ)$. Fix an orientation for the edges $e$ of $Y$ and a spanning tree $\tau$ of $Y$. There is a unique path in $\tau$ (without backtracking) the starts at $y_0$ and ends at the origin $o(e)$ of a given oriented $e \in E(Y)\backslash E(\tau)$, where $E(Y)$, $E(\tau)$ are the edge sets of a $Y$ and $\tau$, respectively.
Continue this path traversing $e$ and then back to $y_0$ along edges of $\tau$, again the last is unique. In this way the oriented edge $e \in E(Y) \backslash E(\tau)$ determines a closed path in $Y$ from $y_0$ to $y_0$.
These $|E(Y) \backslash E(\tau)| = k$ paths generate $\pi_1(Y, y_0)$ freely.
Their images in $H_1(Y, \ZZ)$ generate $H_1(Y, \ZZ)$ as a free $\ZZ$-module. These can be realized by the $k$-closed cycles in $Y$ which start at $t(e)$ and go to $o(e)$ along the unique path in $\tau$ and close up by going from $o(e)$ to $t(e)$ along $e$. These cycles $c_e$ for $e\in E(Y)\backslash E(\tau)$ are a $\ZZ$-basis for $H_1(Y,\ZZ)$. With this we can identify the torus $T(Y)$ with $z \in (S^1)^k$ by setting 
\begin{equation}\label{eqn:twistangles}
\chi_z(c_e) = z_e, \mbox{ for }  z = (z_e)_{e \in E(Y)\backslash E(\tau)}, |z_e| = 1.
\end{equation}

Using the standard basis for functions on $Y$ so that the adjacency matrix of $Y$ is the $n \times n$ matrix whose $v,w$ entry is $1$ if $(v,w) \in E(Y)$ and $0$ otherwise,we find that the matrix $\ad_\chi$ acting on $V_\chi$ has $v,w$ entry
\begin{eqnarray}\label{eqn:BlochMatentries}
a_\chi (v,w) &=& \begin{cases} 0 \mbox{ if there is no edge from } v \mbox{ to } w \mbox{ in } y\\
		           1 \mbox{ if } v \mbox{ is joined to } w \mbox{ in } \tau\\
		             z_e \mbox{ if } o(e) = v, t(e) = w, e \notin E(\tau) \\
		              z_e^{-1} = \bar{z}_e \mbox{ if } o(e)  = w, t(e) = v, e \notin E(\tau)
		  \end{cases} 
\end{eqnarray}
Clearly $\ad_\chi$ is $n\times n$ Hermitian in this basis, as seen explicitly in the following example.

\begin{figure}[h]
	\begin{center}
		\includegraphics[width=0.50\textwidth]{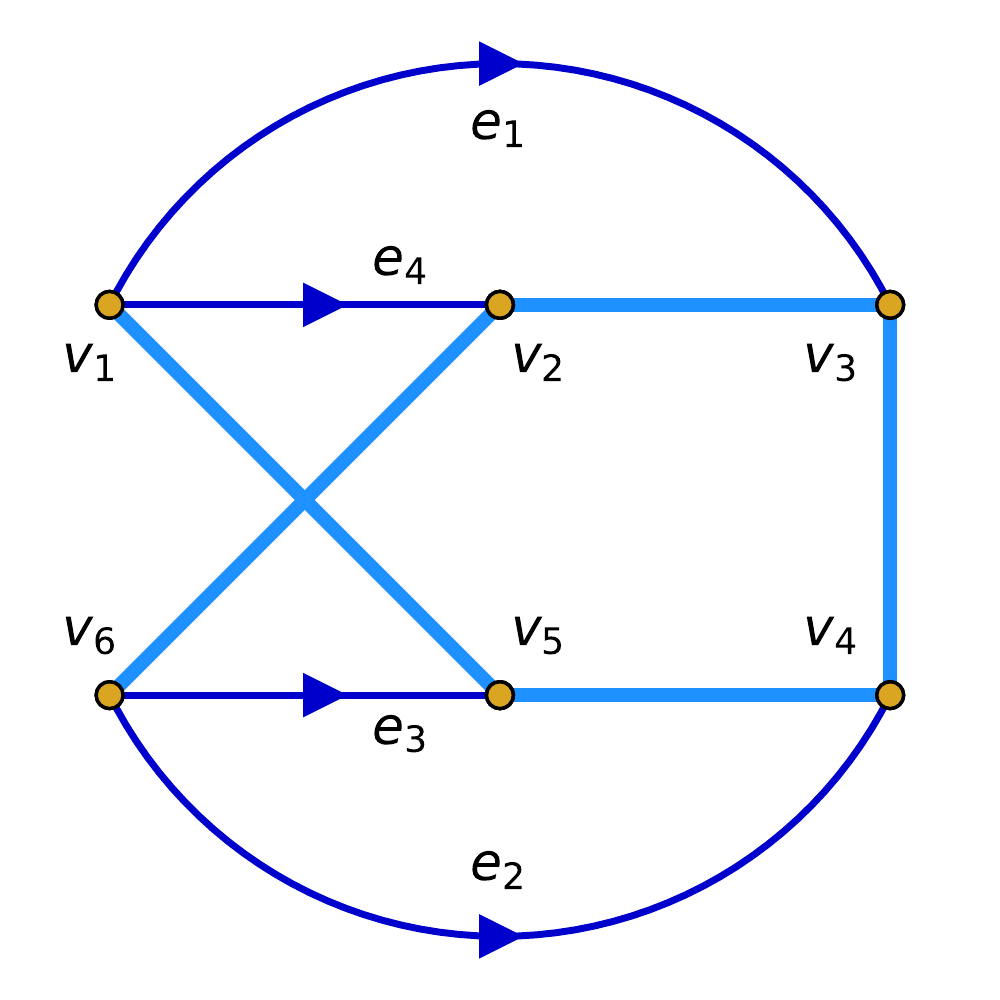}
	\end{center}
	\vspace{-0.6cm}
	\caption{\label{fig:Example1} 
     Example $3$-regular graph $Y$ with $6$ vertices and $9$ edges.  The spanning tree $\tau$ is indicated by the thicker light blue edges.
     The edges $e_1,\cdots, e_4$ not in the spanning tree may freely be redirected to link different decks of a cover of $Y$. If, however, edges of the spanning tree are also redirected, then vertices can become isolated from their own deck, and the cover can become disconnected. The torus $T$ is thus four dimensional, even though $Y$ has $9$ edges.
    } 
\end{figure}

Consider the $3$-regular graph $Y$ shown in Fig. \ref{fig:Example1}. It has $n=6$ vertices,$m=9$ edges, and $k = 4$ edges not in the spanning tree $\tau$, as indicated in Fig. \ref{fig:Example1}. The matrix of $
\ad_\chi$ in the standard basis is:
\begin{equation}\label{eqn:fullBlochMat}
\ad_{\vec{z}} =
 \begin{bmatrix}
    0 & z_4 & z_1 &0 &1& 0 \\
   z_4^{-1} & 0 & 1 & 0 & 0 & 1\\
   z_1^{-1} & 1 & 0 & 1 & 0 & 0 \\
   0 & 0 & 1 & 0 & 1 & z_2^{-1}\\
   1 & 0 & 0 & 1 & 0 & z_3^{-1}\\
   0 & 1 & 0 & z_1 & z_3 & 0
\end{bmatrix}
\end{equation}

The algebraic functions $\lambda_1(\vec{z}), \cdots, \lambda_n(\vec{z})$ can be computed from the secular equation 
\begin{equation}\label{eqn:chareq}
P(\lambda, z) = \det (\lambda I_{n\times n} - A_{\vec{z}}) = 0.
\end{equation}
$P$ is a polynomial in $\lambda$ of degree $n$ with coefficients which are Laurent polynomials of degree one in each variable $z_j$. When we  restrict $P$ to a subtorus $S$ of $T$ as we do in Section \ref{sec:examples}, the number of variables goes down but the degree of those variables in the coefficients goes up. Our connected subtorus $S$ will be chosen to be of dimension $1$ or $2$. In Example $1$, passing to the one-dimensional torus: $z_3 = z_4 = 1$ and $z_1 = z_2$ yields one of our extremal spectral intervals. Note that when $S$ is one dimensional, the corresponding cover of $Y$ is infinite cyclic.

A rare feature shared by our extremal examples, and which is often responsible for special properties, is the existence of flat bands. This occurs if the restriction of $P$ to $S$ (in terms of the $z_j$ variables) has one of the $\lambda_j = \hat{\lambda}$ which is constant on $S$. Equivalently 
\begin{equation}\label{eqn:fbcriterion1}
P(\hat{\lambda},\vec{z}) = 0 \mbox{ for } \vec{z} \in S.
\end{equation}
If there are exactly $r$ $\lambda_j$'s equal to $\hat{\lambda}$ for $\vec{z} \in S$, then we say that the flat band has multiplicity $r$. In terms of the secular polynomial this is equivalent to 
\begin{equation}\label{eqnfbcriterion2}
P (\hat{\lambda}, \vec{z}) = \frac{\partial P}{\partial \lambda} (\hat{\lambda},\vec{z}) = \cdots = \frac{\partial^{r-1} P}{\partial \lambda^{r-1}} (\hat{\lambda},\vec{z}) = 0 \mbox{ for } \vec{z} \in S.
\end{equation}
$\hat{\lambda}$ must be an eigenvalue of $Y$ and the flat bands keep this eigenvalue with very high multiplicity, which works to keep the inclusion in Eqn. \ref{eqn:k0eigs} tight.

The flat bands also have a well-known interpretation in terms of corresponding to localized eigenfunctions of $\ad$ with eigenvalue $\hat{\lambda}$ on the $H_1/S^\perp$ cover $W$ of $Y$ (See \cite{R-S}).
In certain situations, such as the ones in Section \ref{sec:examples}, these localized states are of finite support and their presence can be explained by explicit local configurations.

Our computations of the general $\lambda_j(\vec{z})$ for $\vec{z} \in S$ are numerical and there are certain end points of bands that we need to know exactly. For this analysis we make use of a classical theorem of Rellich \cite{Re} which allows us to conclude that, at least if $S$ is one-dimensional, the $\lambda_j$'s can be chosen to be real analytic functions of $z$ (not just continuous), and the corresponding eigenvectors $v_j$ can also be taken to vary real analytically and also orthogonal to each other.

\section{Construction of Abelian Covers with Large/Extremal Gaps}\label{sec:examples}
\subsection{Bloch-wave Formalism and Generation of Band Spectra}\label{subsec:bloch}

\begin{figure}[h]
	\begin{center}
		\includegraphics[width=1.0\textwidth]{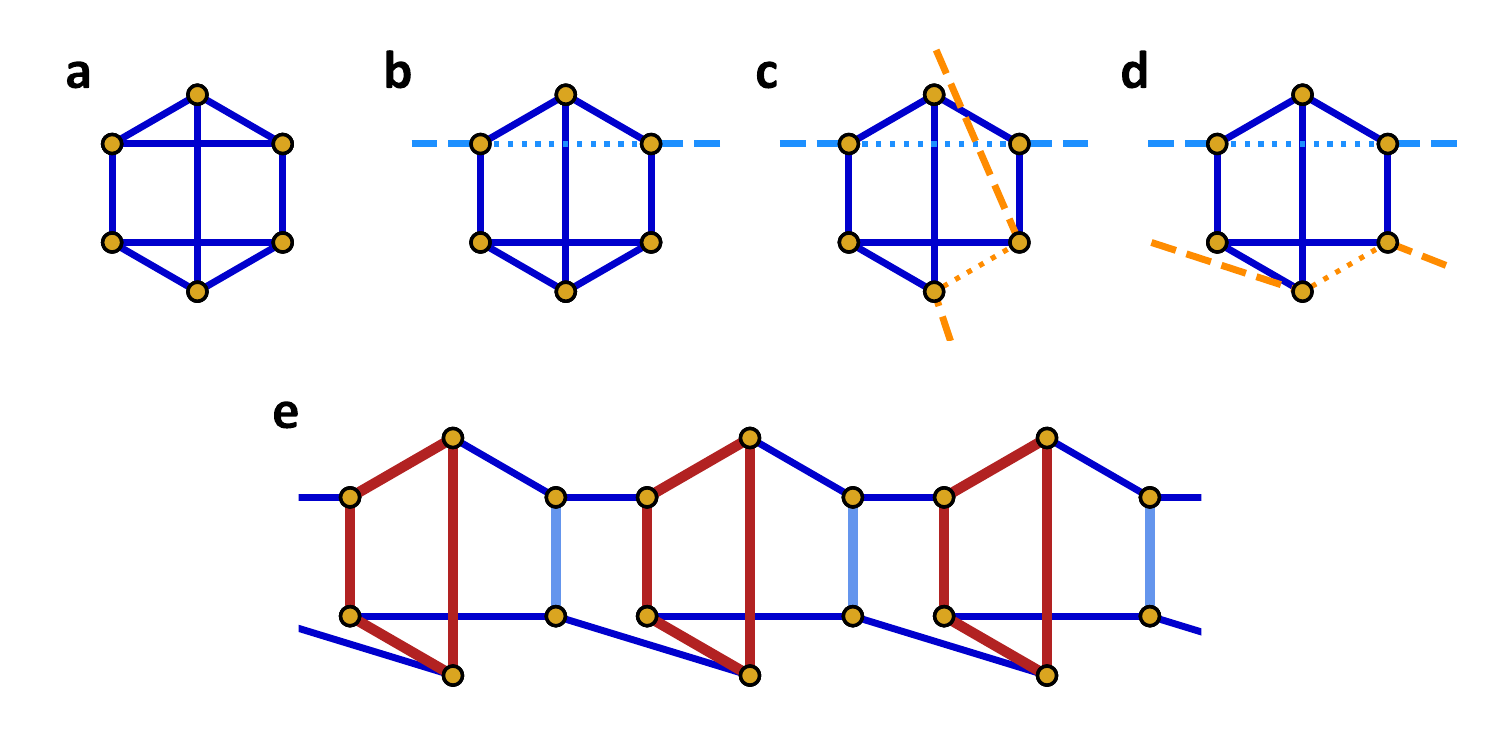}
	\end{center}
	\vspace{-0.6cm}
	\caption{\label{fig:CoverConstruction} 
     \textbf{Cover construction.} \textbf{a} An example seed graph $Y$. \textbf{b} A one-dimensional cover is constructed by severing one link (dotted light blue edge) and reconnecting (dashed light blue edge) to neighboring copies (not shown). \textbf{c} For a two-dimensional cover, two links are severed, one reconnected along $x$ (light blue) and one along $y$ (orange). \textbf{d} A two-link one-dimensional cover is formed by connecting both links in one direction. 
     \textbf{e} Three unit cells of the cover in  \textbf{d}. 
     An alternate drawing of this cover is shown in Fig. \ref{fig:extremals} \textbf{a}\textit{ii}, where the red $4$-cycles become the ``hourglasses'' and the light blue edges the vertical bars.
    } 
\end{figure}

In order to find graphs with large or extremal gap sets, we carried out a targeted numerical search of one- and two-dimensional Abelian covers of small $3$-regular graphs. First, we chose a target seed graph, or unit cell, $Y$ such that $\sigma(Y)$ has large gaps. The book \cite{C-D-S} proved extremely useful for identifying good candidates as it has all $3$-regular graphs up to degree $12$ tabulated, along with their spectra. (Note: the spectra of graphs 3.2 and 3.3 are swapped in this table.) Basic one-dimensional covers $W_{1}(Y)$ can be constructed from $Y$ taking infinitely many copies of $Y$, which we designate by $Y^{(h)}$, and choosing an edge $\tilde{e} \in E(Y)$ through which to connect the $Y^{(h)}$. The edge $\tilde{e}$ has a copy $\tilde{e}_h$ in each $Y^{(h)}$. Each $\tilde{e}_h$ is then cut and reattached to the corresponding point in the next unit cell:
\begin{equation}\label{eqn:estitch}
\tilde{e}_h \rightarrow \tilde{e}_h' =  \begin{cases} o(\tilde{e}_h') = o(\tilde{e}_h)  \\
		 t(\tilde{e}_h') = t(\tilde{e}_{h+1}).
		  \end{cases} 
\end{equation}
The result is a periodic infinite (or finite cyclic) graph $W_{1}(Y, \tilde{e})$.  Two-dimensional covers were constructed in an analogous fashion by laying out copies of $Y$ in a grid and selecting two edges along which to connect:
\begin{equation}\label{eqn:estitch2d}
\tilde{e}_{h,k} \rightarrow \tilde{e}_{j,k}' =  \begin{cases} o(\tilde{e}_{h,k}') = o(\tilde{e}_{h,k})  \\
		 t(\tilde{e}_{h,k}') = t(\tilde{e}_{h+1,k})
		  \end{cases} \\
\mbox{ and     }  \bar{e}_{h,k} \rightarrow \bar{e}_{h,k}' =  \begin{cases} o(\bar{e}_{h,k}') = o(\bar{e}_{h,k})  \\
		 t(\bar{e}_{h,k}') = t(\bar{e}_{h,k+1})
		  \end{cases}.
\end{equation}
The resulting graph $W_{2}(Y, \tilde{e}, \bar{e})$ is structured like a square grid. The covers $W_{1}(Y, \tilde{e})$ and $W_{2}(Y, \tilde{e}, \bar{e})$ are the simplest possible covers with one link between copies (decks) per dimension, and can easily be generated by iterating through all $e \in E(Y)$ and all pairs of edges $e_1, e_2 \in E(Y)$. A sketch of this construction is shown in Fig. \ref{fig:CoverConstruction} \textbf{b} for $W_1$ and \textbf{c} for $W_2$.

Because these structures are Abelian covers of a finite unit cell, they are translation invariant and the full adjacency matrix $\ad_W$ commutes with translations in the link directions. Solutions which are also eigenfunctions of these translations will have the same form $v_\theta$ on each $Y^{(h)}$ except for a phase factor $z = \exp ( i \theta)$ and are known as Bloch waves. In this basis the full Hilbert space can be broken into sectors each of fixed $\theta$ in which $\ad_W$ is block diagonal. 
In this basis, solutions are of the form
\begin{equation}\label{eqn:blochansatz}
\psi_{\theta_1, \theta_2} = v_{\theta_1, \theta_2}(Y^{(h,k)}) \times e^{ i (h \theta_1 + k \theta_2)},
\end{equation}
where $v_{\theta_1, \theta_2}$ is a complex-valued function of $Y$ which depends on the two phases $\theta_1$ and $\theta_2$ for two-dimensional covers, or a single phase angle for a one-dimensional cover. Note that while solutions of this type are not technically $\ell^2$-normalizable, for finite-dimensional Abelian covers such as those considered here, they are in the closure of the $\ell^2$ space. We will therefore compute with them directly rather than including an envelope function which decays sufficiently rapidly at infinity and then taking the limit of the width of the envelope going to infinity.
 
The key feature which makes $\sigma(\ad_W)$ easy to compute numerically is that the action of $\ad_W$ on solutions of this form, i.e. the diagonal blocks, can readily be determined (see for example a solid-state physics textbook such as \cite{Gi}). 
There are $m$ equivalence classes of edges of $W$, corresponding to each of the edges of $Y$, and $\psi_{\theta_1, \theta_2}$ obeys this same symmetry. As a result the full problem can be recast as an effective eigenvalue problem on $Y$ with a modified adjacency operator
\begin{equation}\label{eqn:blochreduction}
\ad_W \psi_{\theta_1, \theta_2} = \lambda \psi_{\theta_1, \theta_2} \Leftrightarrow \ad_\chi v_{\theta_1, \theta_2} = \lambda v_{\theta_1, \theta_2},
\end{equation}
where $\ad_\chi$ is the $\chi$-twisted adjacency operator on $Y$ defined in Eqn. \ref{eqn:BlochMatentries} on a subtorus with two nonzero $z_i$. This definition absorbs the phase factor $e^{ i (h \theta_1 + k \theta_2)}$ into the adjacency operator and replaces the infinite-dimensional Hilbert space of $W$ with infinitely many finite ($n$) dimensional  Hilbert spaces parametrized by the twists angles. The spectrum of $\ad_W$ can then be computed by numerical diagonalization of $\ad_\chi$
in a grid covering the subtorus $S$.
For each value of $\theta_1$ and $\theta_2$, there will be $n$ different solutions. These solutions $\lambda_j$ as a function of $\theta$ are the bands of $W$. 

\begin{figure}[h]
	\begin{center}
		\includegraphics[width=1.0\textwidth]{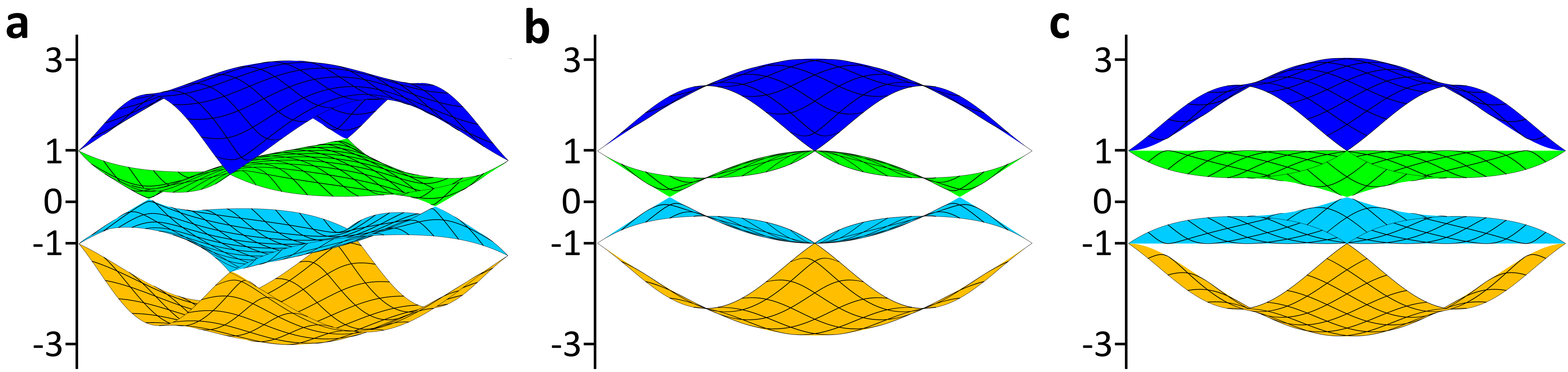}
	\end{center}
	\vspace{-0.3cm}
	\caption{\label{fig:2Dbands} 
     \textbf{Two-dimensional band structure.} \textbf{a} Two-dimensional surface plot of the bands $\lambda_j(\theta_1, \theta_2)$ of a two-dimensional Abelian cover of the multigraph in Fig. \ref{fig:extremals} \textbf{b}\textit{i}. The domain is the unfolded torus $\theta_1 \in [-\pi, \pi], \theta_2 \in [-\pi,\pi]$, i.e. the first Brillouin zone. This seed graph has four vertices, so the cover has four bands, shown in gold, cyan, green, and blue.  \textbf{b} Projection of the bands along the $\theta_1 = \theta_2$ direction. \textbf{c} Projection of the bands along the $\theta_1 = -\theta_2$ direction, which reveals that the green and cyan bands are flat along the line $\theta_1 = \theta_2$. While the full two-dimensional cover is not gapped at zero due to the conical band touches between green and cyan bands clearly visible in \textbf{b}. The one-dimensional cover along $\theta_1 = \theta_2$ is and realizes the maximal gap interval $(-1,1)$, as shown in Fig. \ref{fig:extremals} \textbf{b}\textit{i}-\textit{iii}} 
\end{figure}

In order to find the bands numerically, we discretize the torus $S$ in an evenly spaced-$N\times N$ square grid for two-dimensional examples, and size $N$ grid for one-dimensional ones. At each point of the grid we find the matrix $\ad_\chi$ and diagonalize it numerically using a standard numerical diagonalizer optimized for Hermitian matrices (from numpy.linalg python wrapper for BLAS and LAPACK). 
Sample plots of the eigenvalues as surfaces as a  function of $\theta_1$ and $\theta_2$ are shown in Fig. \ref{fig:2Dbands}.
Collecting all of the eigenvalues from all values of $\theta_1$ and $\theta_2$ then provides an estimate of spectrum of $\ad_W$. These eigenvalues are sorted, and we examine all the intervals between adjacent eigenvalues. Most such intervals are spurious and merely represent the discretization of the grid in $\theta$. We therefore reject all intervals smaller than the generous threshold of $0.05$, and the remaining large intervals are interpreted as the gaps of $\ad_W$. Generally, this provides an overestimate of the gaps because the leading source of error comes from possibility that the mesh in $\theta$ does not include the exact maximum or minimum of a band, rather than from the numerical diagonalizer. In the case of gaps bordered by flat bands, this step size is not an issue, and the numerical gap will typically be an underestimate which is limited by errors from the numerical diagonalizer around the $10^{-14}$ level.

When generating covers to compute, we neglect to identify a spanning tree and instead compute the spectrum of all possible $W_{2}(Y, \tilde{e}, \bar{e})$, 
all of which will be connected since $Y$ is $3$-regular and we are redirecting at most two edges.
Many of the resulting covers are redundant if $Y$ has a high degree of symmetry, which is the case for the cells shown in Fig. \ref{fig:extremals} \textbf{a}\textit{i} and \textbf{b}\textit{i}. While some $W_{2}$ with large gap intervals were found, none realized maximal gap intervals. However, each two-dimensional torus contains infinitely many subtori (circles in this case), corresponding to setting a relation
\begin{equation}\label{eqn:abtheta}
a \theta_1 = b \theta_2,
\end{equation}
for integer coefficients $a$ and $b$. Such a relation corresponds to a more complicated one-dimensional cover in which multiple links go between unit cells (decks). For example, setting $\theta_1 = \theta_2$ corresponds to cutting two edges in $Y$ and connecting them both to the same neighboring unit cell, rather than in two separate directions of a grid. Fig. \ref{fig:CoverConstruction} \textbf{d},\textbf{e} shows one such cover.
The spectrum and bands of each of these two-link one-dimensional covers can be found by looking along the corresponding line in the full two-dimensional solution.  In this manner, the square-grid two-dimensional covers $W_2$ were used to search the space of two-link one-dimensional covers, beyond the simple one-link cases realized for $W_1 = W_2 (\theta_2 = 0)$.


Using this search method, two sets of special one-dimensional covers were found. First, four non-planar examples that realized the extremal gap intervals $(-2,0)$ and $(-1,1)$ and are described in Section \ref{subsec:extremalgaps}. Second, a set of four planar examples with large (but not maximal) gaps, the union of which cover the interval $[-2,0]$. As shown in Section \ref{sec:T}, with these four graphs as input, the map $\TT$ can produce a gap anywhere in the interval $[-3,3)$.

\subsection{Extremal Gap Sets}\label{subsec:extremalgaps}
\begin{figure}[h]
	\begin{center}
		\includegraphics[width=0.95\textwidth]{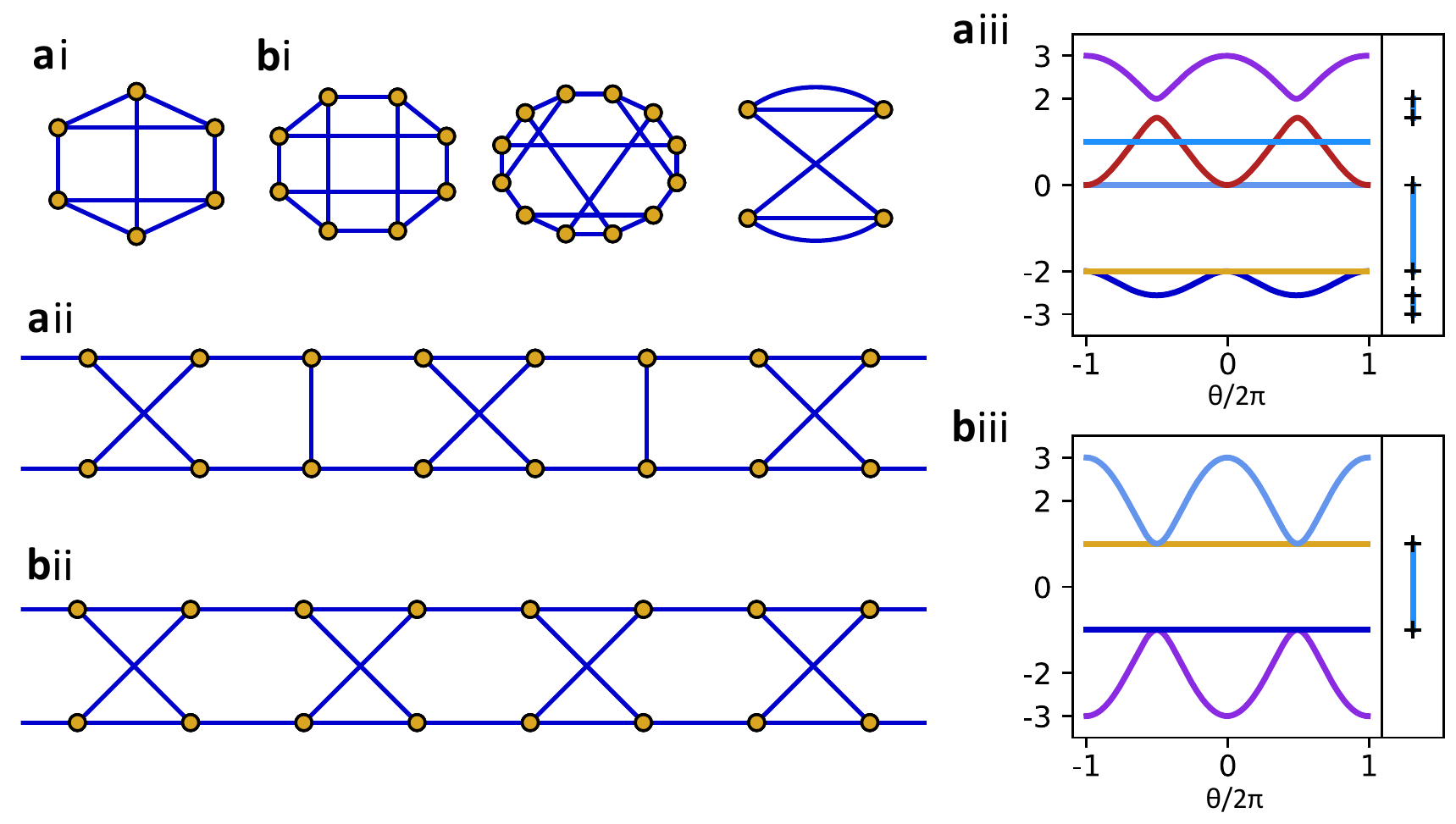}
	\end{center}
	\vspace{-0.3cm}
	\caption{\label{fig:extremals} 
    \textbf{Graphs with extremal gap intervals}. 
    \textbf{a}\textit{i} - \textbf{b}\textit{i} Four examples of graphs $Y$ carrying infinite Abelian covers realizing extremal gap intervals. The graph in \textbf{a}\textit{i} yields the extremal cover shown in \textbf{a}\textit{ii}, while the three graphs in \textbf{b}\textit{i} yield the same extremal cover, shown in \textbf{b}\textit{ii}. 
    The two seed  in graphs \textbf{b}\textit{i} without multiple edges each contain $2$ and $3$ copies of the fundamental domain of the cover. In order to obtain only one copy per deck, the cover must be made from the multigraph in \textbf{b}\textit{i}.
         \textbf{a}\textit{iii} - \textbf{b}\textit{iii} Numerical Bloch-wave bands for each cover graph as a function of character angle $\theta$. Each plot shows two fundamental domains, i.e. two loops around the dual torus. The gap sets for each example are highlighted in the insets. Solid lines indicate gap intervals  and cross markers their boundaries. Graph \textbf{a} realizes the  extremal gap interval $(-2,0)$, and \textbf{b} realizes $ (-1,1)$.
    } 
\end{figure}


The four initial cells $Y$ which yield Abelian covers which realize maximal gap intervals are shown in Fig. \ref{fig:extremals} \textbf{a}\textit{i} - \textbf{b}\textit{i}. In all four cases the extremal covers were found to be two-link one-dimensional covers and were initially identified as special directions in two-dimensional examples. These extremal covers, $\bar{W}_a$ and $\bar{W}_b$, are shown in Fig. \ref{fig:extremals} \textbf{a}\textit{ii} - \textbf{b}\textit{ii}, and details of the construction of $\bar{W}_a$ are shown in Fig.~\ref{fig:CoverConstruction}. The three graphs in Fig. \ref{fig:extremals} \textbf{b}\textit{i} produce the same extremal graph. The first (cube) graph gives rise to two copies of the fundamental domain per deck and the second results in three. In order to obtain only a single copy of the fundamental domain the starting graph must be the final graph in  \textbf{b}\textit{ii}, which has multiple edges.
The common feature between all of these graphs is a $4$-cycle connected to the left and right at opposite pairs of corners. This feature is a drawn as an hourglass in Fig. \ref{fig:extremals} \textbf{a}\textit{ii} - \textbf{b}\textit{ii} and gives rise to all of the flat bands that these models exhibit.


\begin{figure}[h]
	\begin{center}
		\includegraphics[width=1.0\textwidth]{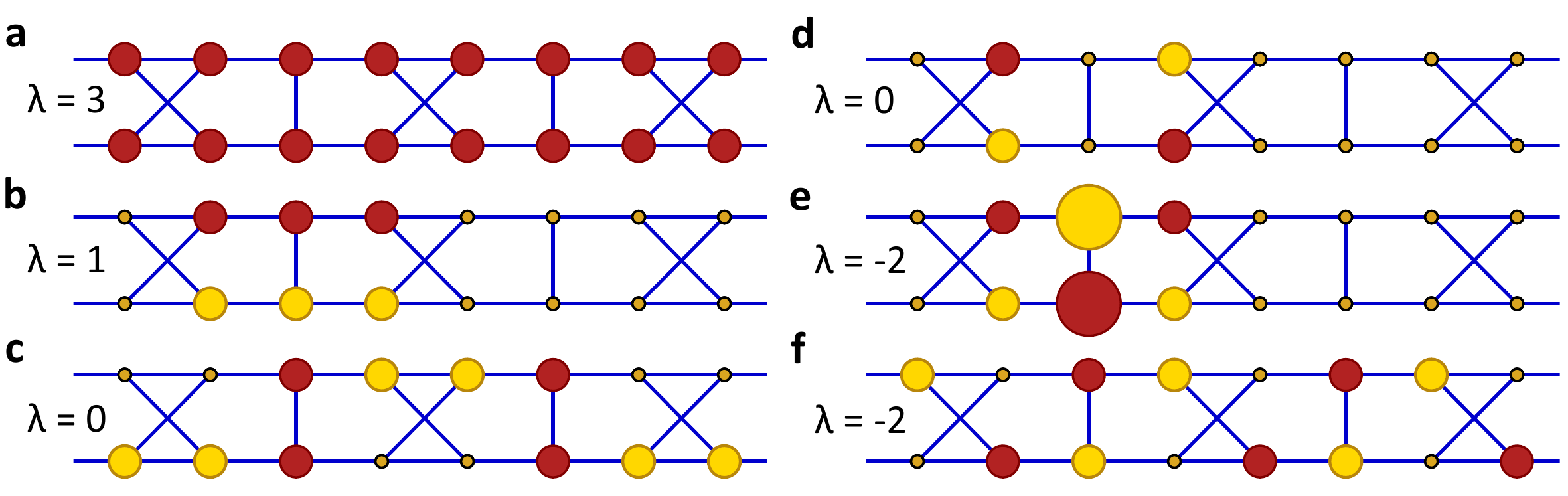}
	\end{center}
	\vspace{-0.3cm}
	\caption{\label{fig:eiginestatesa} 
    \textbf{Eigenstates of graph a}. Eigenvectors of the Abelian cover shown in Fig. \ref{fig:extremals} \textbf{a}\textit{ii} at $\theta_t = 0$, shown in order of bands from largest to smallest lambda. Unlike the numerical spectrum in Fig. \ref{fig:extremals} \textbf{b}\textit{iii}, all eigenvectors and eigenvalues are exact. Eigenvectors are plotted as colored circles overlaid on the vertices of the graph. Red (dark) circles indicate positive sign and yellow (pale) negative. The radius of the circle indicates the amplitude of the eigenvector. Plotted eigenvectors are unnormalized and not orthogonalized in order to represent the graph's structure most simply: except for the state in \textbf{e} which goes up to values of $\pm 2$, a full set of (non-orthogonal) eigenvectors can be produced from vectors with entries only $0, \pm1$. Eigenvectors corresponding to flat bands have compact support. 
    } 
\end{figure}

The band structures of $\bar{W}_a$ and $\bar{W}_b$ are shown in  Fig. \ref{fig:extremals} \textbf{a}\textit{iii} - \textbf{b}\textit{iii}. Each band is color coded, showing the continuous evolution of the eigenvalues $\lambda_j(S)$ versus the twist angle $\theta$, and the gap sets for each are highlighted in insets to the right of the main plot. Numerically, each extremal cover was determined to have gap intervals which are consistent with $(-2,0)$ and $(-1,1)$. In order to prove that these gaps are exact, it is necessary to supplement the numerical band structure calculation and establish two additional facts exactly:
\begin{enumerate}

\item the flat bands are exactly flat and located precisely at $\{-2,0\}$ for $\bar{W}_a$ and  $\{-1,1\}$ for $\bar{W}_a$,

\item the curved bands never cross these flat bands.

\end{enumerate}
In the absence of a fully analytic solution for $\lambda_j(\theta)$, we make use of the fact that the band structure is periodic in $\theta$, along with a theorem by Rellich \cite{Re} that the eigenvalues as function of $\theta$ can be taken to be real analytic, in order to show that these gaps are exact. Both band structures are periodic in $\theta$ with a period of $2 \pi$, and the bands are very well behaved functions of $\theta$. Therefore, in both cases there is only one point in the dual torus where the curved bands approach the flat bands and near which they could potentially cross: $\theta_t = 0$ for $\bar{W}_a$ and $\theta_t =\pi$ for $\bar{W}_b$. 

\begin{figure}[h]
	\begin{center}
		\includegraphics[width=1.0\textwidth]{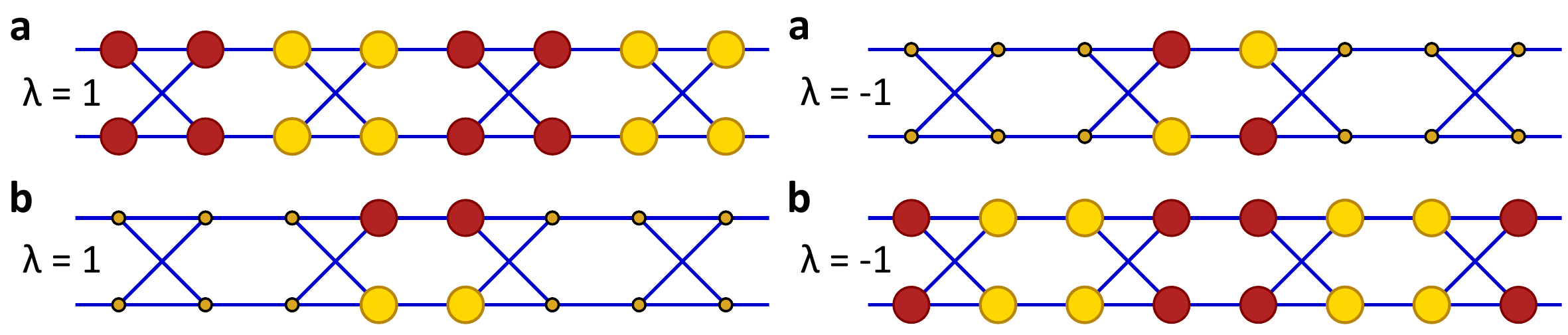}
	\end{center}
	\vspace{-0.3cm}
	\caption{\label{fig:eiginestatesb} 
    \textbf{Eigenstates of graph b}. Exact eigenvectors of the Abelian cover shown in Fig. \ref{fig:extremals} \textbf{b}\textit{ii} at $\theta_t = \pi$, shown in order of bands from largest to smallest lambda. Unlike the numerical spectrum in Fig. \ref{fig:extremals} \textbf{b}\textit{iii}, all eigenvectors and eigenvalues are exact. Eigenvectors are plotted as colored circles overlaid on the vertices of the graph. Red (dark) circles indicate positive sign and yellow (pale) negative. The size of the circle indicates the amplitude of the eigenvector. Plotted eigenvectors are unnormalized and not orthogonalized in order to represent the graph's structure most simply. Eigenstates corresponding to flat bands have compact support.
    } 
\end{figure}

To characterize this point, we examine the twisted adjacency operator $\ad_{\theta}$ at the band touch point. While $\ad_\theta$ is in general Hermitian, at $\theta = \{0,\pi\}$ it is a real symmetric matrix with entries either $-1$, $0$, or $1$, and admits real eigenvectors. Due to this particularly simple structure, the eigenvectors and eigenvalues can be found exactly, and using Eqn. \ref{eqn:blochansatz} these can then be converted into eigenvectors of $\ad_W$. In the case of the dispersive (non-flat) bands, the only exact eigenvectors are of this Bloch-wave form. However, the degeneracy of the flat bands allows other bases to be chosen in which the eigenvectors are localized. In this case, the Bloch-waves can be understood as a sum of translates of these localized solutions each with a phase twist $\exp(i h\theta)$. The resulting eigenvectors are shown in Fig. \ref{fig:eiginestatesa} for $\bar{W}_a$ and Fig. \ref{fig:eiginestatesb} for $\bar{W}_b$. The eigenvectors are plotted as colored circles overlaid on the sites of the graph where the size of the circle indicates the magnitude of the vector on that site and the color the sign. 

The action of $\ad_W$ on these vectors can then be double checked by adding up the value on all neighboring sites and comparing to the on-site value. This therefore establishes that at the band touch angle $\theta_t$ the cover $\bar{W}_a$ has eigenvalues $\{3,1,0,0,-2,-2 \}$ and $\bar{W}_b$ has eigenvalues $\{1,1,-1,-1\}$, exactly. Note that we have chosen to draw the eigenvectors so that they are as simple as possible and the eigenvalues easiest to verify. They are unnormalized and may be only linearly independent rather than orthogonal. Proper orthonormal eigenbases can be found via Graham-Schmidt on finite cyclic covers and extrapolated to infinite ones. 
(However, the resulting states are needlessly difficult to visualize.) In the case of the five flat bands, we have plotted states in the localized basis where they are particularly simple and of compact support. Translates of these states are also eigenstates with the same eigenvalue. These can then be plugged in Eqn. \ref{eqn:blochansatz} to produce Bloch-wave solutions as a function of $\theta$, which will in turn all have the same eigenvalue. Thus, these bands are exactly flat, and not merely numerically so.

Finally, it only remains to establish that the dispersive bands (which are analytic by Rellich \cite{Re}) do not encroach on the gap by crossing the flat bands. In both cases, $\ad_\theta$ is symmetric around $\theta_t$ such that 
\begin{equation}\label{eqn:Hsymmetry}
\ad_{\theta_t +\delta} = \mbox{transpose}(\ad_{\theta_t - \delta}).
\end{equation}
and by symmetry, the entire band structure $\{\lambda_j \}$ must be symmetric about $\theta_t$. Since the flat bands are already symmetric and the dispersive bands never cross, this forces all the bands in $\sigma(\bar{W}_a)$ and $\sigma(\bar{W}_b)$ to be symmetric about $\theta_t$ individually and constrains the first derivative versus $\theta$ of each band to zero at $\theta_t$. The numerical calculations already establish that the dispersive bands have non-zero second derivative at $\theta_t$, and it then follows that $\theta_t$ is a local (and in fact global) extremum of all the dispersive bands of both $\bar{W}_a$ and $\bar{W_b}$. Hence, these graphs realize the gap intervals $(-2,0)$ and $(-1,1)$ exactly.

The spectrum of $\bar{W}_b$ follows trivially from here since it has no other gaps. $\bar{W}_a$ has two other gaps whose extrema are at $\theta = \pi$. An analogous treatment can be carried out here with slightly more algebra required to determine the eigenvectors and eigenvalues. The exact spectra and gaps are
  \begin{eqnarray}\label{eqn:exampleExactGapsa}
\sigma(\bar{W}_a) & =&  \left[ -\frac{1+ \sqrt{17}}{2}, -2 \right] \bigcup \left[0, \frac{-1 + \sqrt{17}}{2}\right] \bigcup \ [2,3]   \\\ [-3,3] \backslash  \sigma(\bar{W}_a)  & =&  \left[-3,-\frac{1+ \sqrt{17}}{2} \right) \bigcup \ (-2,0)\  \bigcup \left(   \frac{-1 + \sqrt{17}}{2}, 2   \right),   \nonumber
 \end{eqnarray}
 and
  \begin{eqnarray}\label{eqn:exampleExactGapsb}
\sigma(\bar{W}_b) & =&  \left[ -3,-1 \right] \bigcup \left[1,3 \right]   \\
\ [-3,3] \backslash  \sigma(\bar{W}_b)  & =& (-1,1).    \nonumber
 \end{eqnarray}

The infinite cubic graphs $\bar{W}_a$ and $\bar{W}_b$ can be used to construct infinite sequences of finite graphs whose spectra are contained in $\sigma(\bar{W}_a)$ and $\sigma(\bar{W}_b)$ by taking suitable quotients. For $\alpha = a$ or $b$, let $G_\alpha$ be the automorphism group of $\bar{W}_\alpha$, and let $\Gamma_\alpha$ be a subgroup of $G_\alpha$, then the spectrum of $\ad$ restricted to the $\Gamma_\alpha$ periodic functions on $\bar{W}_\alpha$ is contained in $\sigma(\bar{W}_\alpha)$. This follows from $G_\alpha$ being amenable. If $\Gamma_\alpha$ acts freely on the vertices of $\bar{W}_\alpha$, i.e. any element $\gamma \neq 1$ in $\Gamma_\alpha$ fixes none of the vertices of $\bar{W}_\alpha$, then the quotient $\bar{W}_\alpha / \Gamma_\alpha$ is a multigraph whose spectrum is contained in $\sigma(\bar{W}_\alpha)$. If $\Gamma_\alpha$ acts without fixing any edges, then the quotient is a graph. We examine each case $\alpha = a, b$ in turn.

\begin{figure}[h]
	\begin{center}
		\includegraphics[width=1.0\textwidth]{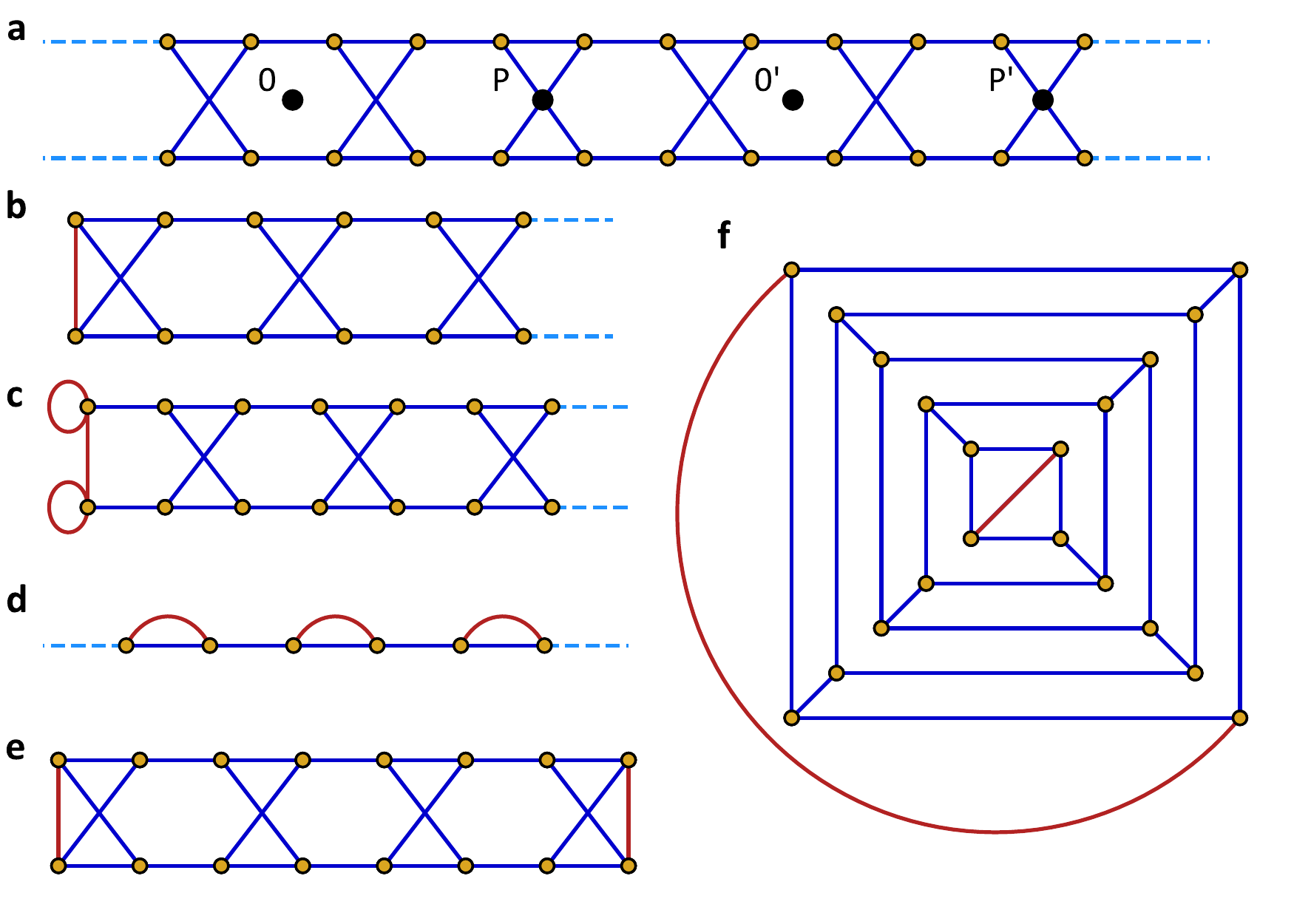}
	\end{center}
	\vspace{-0.6cm}
	\caption{\label{fig:capping} 
    \textbf{Finite planar quotients of $\bar{W}_b$}. \textbf{a} The infinite graph $\bar{W}_b$. Four sample involution symmetry points are indicated by black dots. \textbf{b} The quotient obtained with respect to the automorphism $\sigma_0$: rotation about $O$ or $O'$ by $\pi$. New edges induced by the quotient are indicated in red. In this case, no loops or multiple edges appear. \textbf{c} The quotient with respect to $\sigma_P$. In this case, two loops appear. \textbf{d} The quotient with respect to reflection about the central axis. Infinitely many multiple edges appear. \textbf{e}, \textbf{f} The quotient with respect to $\sigma_O$ and $\sigma_{O'}$, when $O$ and $O'$ are four unit cells apart. This quotient is a planar graph which is $(-1,1)$ gapped.
    } 
\end{figure}

\begin{figure}[h]
	\begin{center}
		\includegraphics[width=1.0\textwidth]{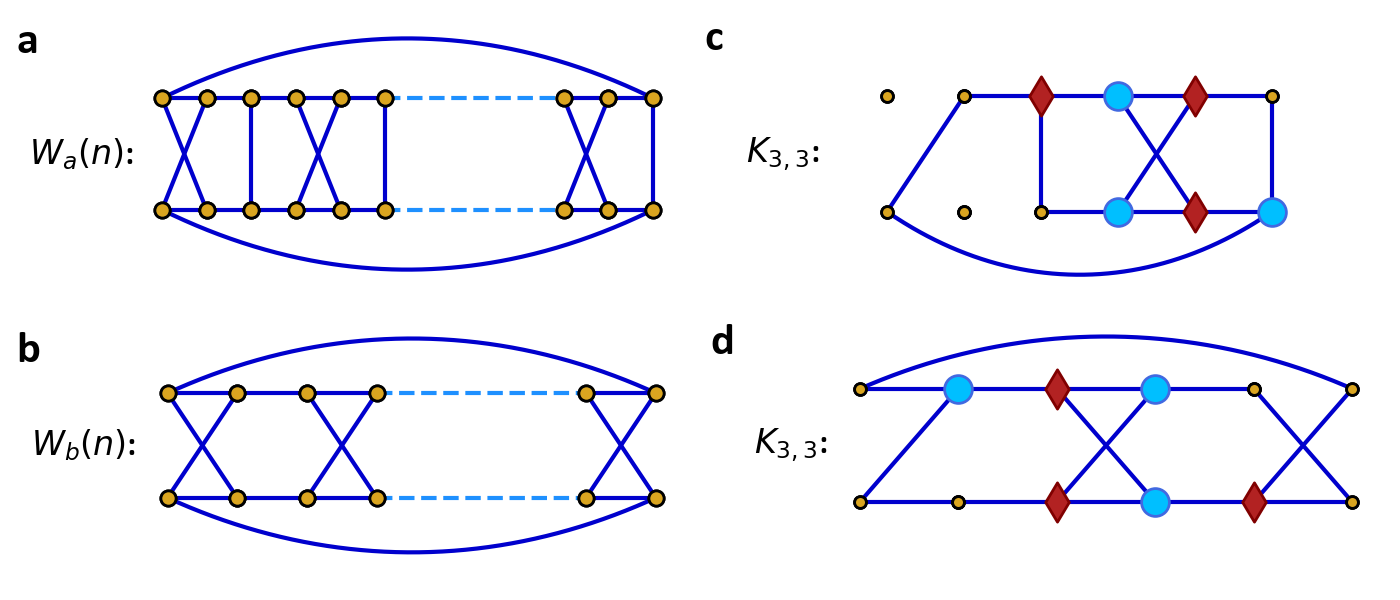}
	\end{center}
	\vspace{-0.6cm}
	\caption{\label{fig:Hamburger} 
    \textbf{Non-planar quotients}. \textbf{a} Finite periodic quotient $W_a(n)$ of the graph $\bar{W}_a$. \textbf{b} Finite periodic quotient $W_b(n)$ of the graph $\bar{W}_b$. \textbf{c},\textbf{d} Sketches showing that $W_a(2)$ and $W_b(3)$ contain subgraphs which are topologically equivalent to the complete bipartite graph on two sets of three vertices:  $K_{3,3}$. One set of vertices is indicated by the light-blue circles, and one by 
    33red diamonds. The existence of this type of subgraph shows that $W_a(n)$ and $W_b(n)$ are non-planar for $n \geq2$ and $n\geq3$, respectively.
    } 
\end{figure}

Consider first $\bar{W}_b$. Its automorphism group is generated by four types of elements.
\begin{itemize}
    \item [(i)] Translations $t(n)$ by n unit cells. The quotients $\bar{W}_b / \left< t(n) \right>$ for $n \geq 2$ are the hamburger graphs $W_b(n)$ shown in Fig. \ref{fig:Hamburger} \textbf{b}.
    
    \item[(ii)] The involution $\sigma_O$ rotating about a central point $O$ by $\pi$. Two example points $O$ and $O'$ are shown in Fig. \ref{fig:capping} \textbf{a}. The quotient $\bar{W}_b / \left< \sigma_O \right> $ is the graph shown in Fig. \ref{fig:capping} \textbf{b}.
    
    \item[(iii)] The involution $\sigma_P$ rotating about a central point $P$ by $\pi$. Two example points $P$ and $P'$ are shown in Fig. \ref{fig:capping} \textbf{a}. The quotient $\bar{W}_b / \left< \sigma_P \right> $ is a multigraph, shown in Fig. \ref{fig:capping} \textbf{c}.
    
    \item[(iv)] The reflection $\mathcal{R}$ about the central axis, which switches the top and bottom vertices. $\bar{W}_b / \left< \mathcal{R} \right> $ is the multigraph shown in Fig. \ref{fig:capping} \textbf{d}.
    
\end{itemize}

The hamburger graphs $W_b(n)$ (shown in Fig. \ref{fig:Hamburger} \textbf{b}) have $4n$ vertices. They are bipartite, and $\sigma(W_b(n)) \subset \sigma(\bar{W}_b)$. Hence, $W_b(n)$ is $(-1,1)$ gapped. These $W_b(n)$'s were previously constructed and shown directly to be $(-1,1)$ gapped in Ref. \cite{G-M}. The graph $W_b(2)$ is the cube and is planar, but for $n \geq3$ $W_n(n)$ is not planar. By Kuratowski's theorem \cite{Kuratowski} the only obstruction for a cubic graph to be planar is that it contain a topological $K_{3,3}$. Such a $K_{3,3}$ is shown in Fig. \ref{fig:Hamburger} \textbf{d} for $W_b(3)$, and the same is true for $W_b(n)$ with $n > 3$.

To produce finite planar quotients of $\bar{W}_b$, we use two involutions of type (ii) centered at two distinct point $O$ and $O'$ which are $n$ unit cells apart. The quotient graphs $P_b(n) := \bar{W}_b / \left< \sigma_O ,\sigma_{O'} \right>$ have size $4n$ and are planar with four triangular faces and $2(n-1)$ hexagonal faces. The resulting quotient for $n= 4$ is shown in Fig. \ref{fig:capping} \textbf{e}, in the realization that derives naturally from \textbf{a}. Fig. \ref{fig:capping} \textbf{f} shows an alternate realization for $n = 5$ which is manifestly planar. The planar graphs $P_b(n)$ are $(-1,1)$ gapped, proving the corresponding statement in Theorem \ref{thm:extremalgaps}.

\begin{figure}[h]
	\begin{center}
		\includegraphics[width=1.0\textwidth]{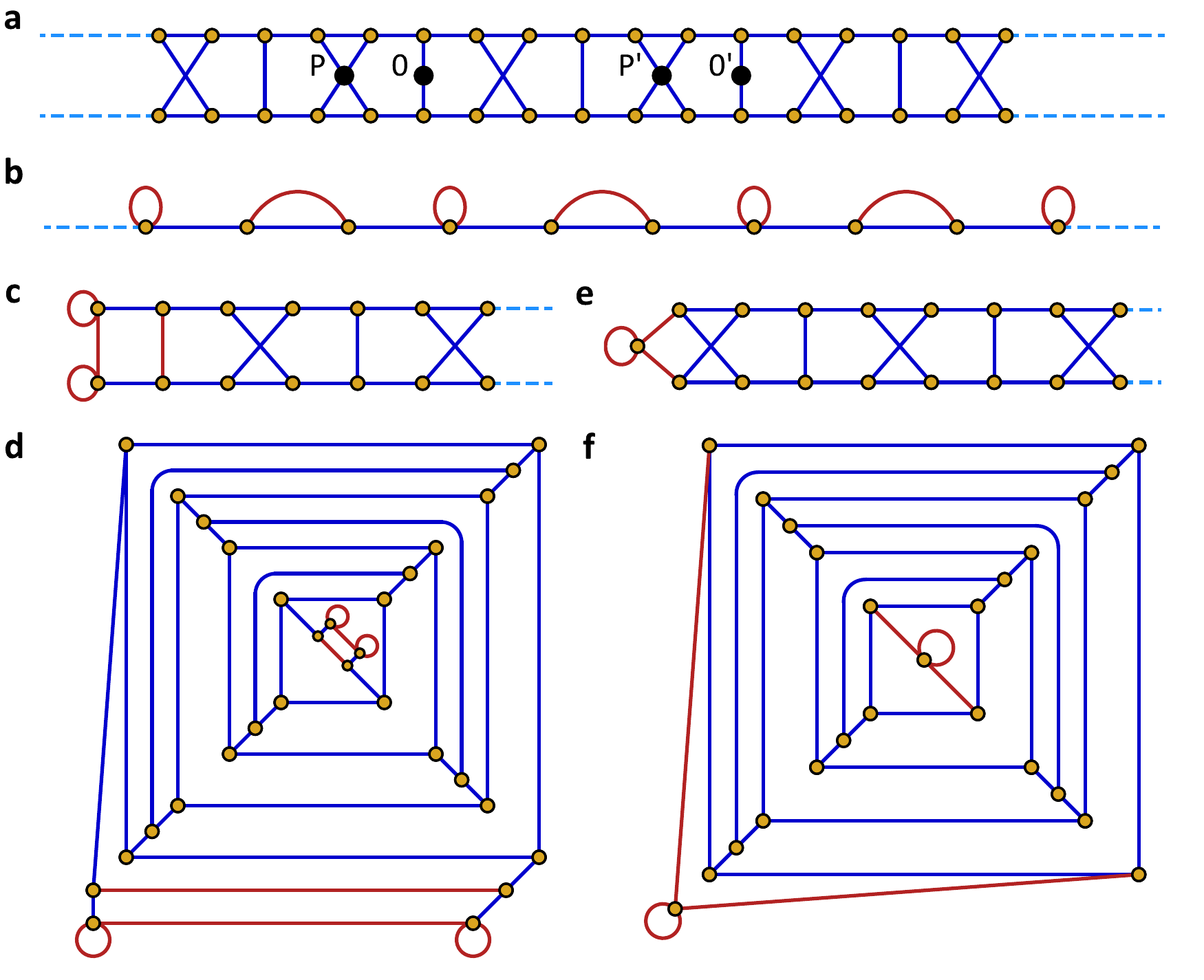}
	\end{center}
	\vspace{-0.6cm}
	\caption{\label{fig:capping2} 
    \textbf{Finite planar quotients of $\bar{W}_a$}. \textbf{a} The infinite graph $\bar{W}_a$. Four sample involution symmetry points are indicated by black dots. \textbf{b} The quotient with respect to reflection about the central axis. New edges induced by the quotient are indicated in red. Infinitely many loops appear.
    \textbf{c} The quotient obtained with respect to the automorphism $\sigma_0$: rotation about $O$ by $\pi$. New edges induced by the quotient are indicated in red. In this case, two loops appear. If a quotient is taken with respect to two distinct $\sigma_O$ and $\sigma_{O'}$, the result is a planar $(-2,0)$ gapped multigraph with four loops, shown in \textbf{d}. \textbf{e} The quotient with respect to $\sigma_P$. In this case, one loop appears. If a quotient is taken with respect to two distinct $\sigma_P$ and $\sigma_{P'}$, the result is a planar $(-2,0)$ gapped multigraph with two loops, shown in \textbf{d}.
    }
\end{figure}

Next we turn to $\bar{W}_a$. Its automorphism group is also generated by four types of elements.
\begin{itemize}
    \item [(i)] Translations $t(n)$ by n unit cells. The basic unit cell is now larger, consisting of an hourglass and a vertical bar. The quotients $\bar{W}_a / \left< t(n) \right>$ for $n \geq 1$ are the hamburger graphs $W_a(n)$ shown in Fig. \ref{fig:Hamburger} \textbf{a}.
    
    \item[(ii)] The involution $\sigma_O$ rotating about a central point $O$ by $\pi$. Two example points $O$ and $O'$ are shown in Fig. \ref{fig:capping2} \textbf{a}. The quotient $\bar{W}_a / \left< \sigma_O \right> $ is the multigraph shown in Fig. \ref{fig:capping2} \textbf{e}.
    
    \item[(iii)] The involution $\sigma_P$ rotating about a central point $P$ by $\pi$. Two example points $P$ and $P'$ are shown in Fig. \ref{fig:capping2} \textbf{a}. The quotient $\bar{W}_a / \left< \sigma_P \right> $ is also a multigraph, shown in Fig. \ref{fig:capping2} \textbf{c}.
    
    \item[(iv)] The reflection $\mathcal{R}$ about the central axis, which switches the top and bottom vertices. $\bar{W}_a / \left< \mathcal{R} \right> $ is the multigraph shown in Fig. \ref{fig:capping2} \textbf{b}.
    
\end{itemize}

The hamburger graphs $W_a(n)$ (shown in Fig. \ref{fig:Hamburger} \textbf{a}) have $6n$ vertices, and $\sigma(W_a(n)) \subset \sigma(\bar{W}_a)$. In particular, $W_a(n)$ is $(-2,0)$ gapped, which establishes the corresponding claim in Theorem \ref{thm:extremalgaps}.
As with the $W_b$'s, $W_a(1)$ is planar (see Fig. \ref{fig:extremals} \textbf{a}\textit{i}), while $W_a(n)$, $n \geq 2$ are not. The topological $K_{3,3}$'s that these contain are shown in Fig. \ref{fig:Hamburger} \textbf{c}.

None of the quotient graphs of $\bar{W}_a$ are planar, and we do not know if $(-2,0)$ can be planar gapped. However, if we allow multigraphs, then this can be done. Chosing two involutions $\sigma_O$ and $\sigma_{O'}$ which are $n$ unit cells apart yields a multigraph $P_a(n) := \bar{W}_a / \left<\sigma_O, \sigma_{O'} \right>$, which looks like Fig. \ref{fig:capping2} \textbf{e} at both ends. A manifestly planar realization of this quotient with $n=4$ is shown in Fig. \ref{fig:capping2} \textbf{f}. The graphs $P_a(n)$ are planar multigraphs with two loops, and $|P_a(n)| = 6n$.

If instead we chose two involutions $\sigma_P$ and $\sigma_{P'}$ of type (ii) with $P$ and $P'$ which are $n$ unit cells apart, we obtains the mutigraphs $Q_a(N) : = \bar{W}_a \backslash \left<\sigma_P, \sigma_{P'} \right>$, which looks like Fig. \ref{fig:capping2} \textbf{c} at both ends. A manifestly planar realization of this quotient with $n=4$ is shown in Fig. \ref{fig:capping2} \textbf{d}. The graphs $Q_a(n)$ are planar multigraphs with four loops, and $|Q_a(n)| = 6n$.

The spectra of both $P_a(n)$ and $Q_a(n)$ are contained in $\sigma(\bar{W}_a)$, and hence, these are planar multigraphs which are $(-2,0)$ gapped, proving the corresponding statement in Theorem \ref{thm:extremalgaps}.

In forthcoming work with Fan Wei, we construct planar multigraphs with are $(-2,0)$ gapped and have exactly two multiple edges and no loops. They are not realized as quotients of $\bar{W}_a$, but rather as two-sided ``cappings'' of it: a construction and analysis that we develop in order to study the gap sets for fullerene graphs.

\subsection{Planar Gap Sets}\label{subsec:planargaps}
\begin{figure}[h]
	\begin{center}
		\includegraphics[width=0.95\textwidth]{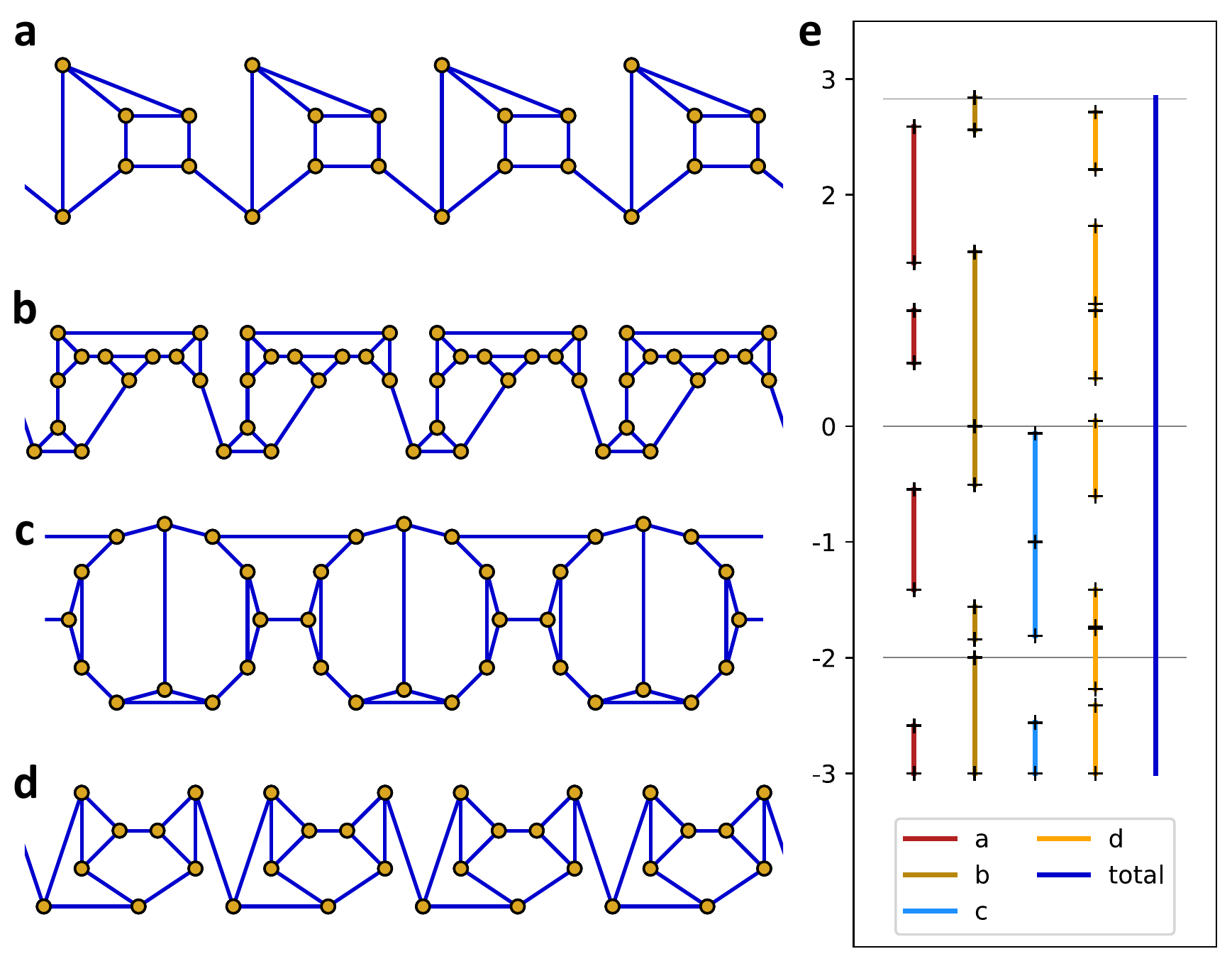}
	\end{center}
	\vspace{-0.3cm}
	\caption{\label{fig:planarexhaustion} 
    \textbf{Planar gaps covering} $[-2,0]$. 
    \textbf{a}-\textbf{d} Finite sections of four infinite planar graphs whose finite periodic quotients are also planar and which were obtained as one-dimensional Abelian covers of small $3$-regular graphs. The gap sets of these four graphs are shown in \textbf{e} in alphabetical order from left to right. Solid lines are used to indicate the gap intervals, and cross marks the limits of the gap intervals. Horizontal grey lines are guides to the eye which indicate the special values $-2$, $0$, and $2 \sqrt{2}$. Except for the special point $\lambda = -3$, the ends of all the gap intervals are open and not included. For example, graph \textbf{c} has a flat band at $\lambda = -1$, and so this point is only gapped in graph \textbf{a}. The union $\mathbb{G}$ of all four of these gaps sets, shown in dark blue, includes not only the interval $[-2,0]$ needed to establish Theorem \ref{thm:planargaps}, but also $[-3, 2\sqrt{2}]$.
    } 
\end{figure}



While most of the covers generated above (and all of the extremal covers) are non-planar, special sets of links generate planar Abelian covers. 
In the case of one-dimensional planar covers, all finite cyclic versions of the cover are also planar. This can be seen by drawing the unit cells in an annular geometry, instead of a straight ribbon. 
Four such examples which yield relatively large gap sets are shown in Fig. \ref{fig:planarexhaustion} \textbf{a}-\textbf{d}, and their gap sets are shown in Fig. \ref{fig:planarexhaustion} \textbf{e}. We denote these four example graphs by $P_a , \cdots , P_d$ and their gap sets by $\GG(P_a), \cdots \GG(P_d)$. 
The gap sets are found by taking the complement of numerical estimates of the bands. The exact edges of the bands and gaps are therefore uncertain, primarily due to discretization in $\theta$ when computing the bands. However, these four graphs were chosen such that their gaps overlap by more than the numerical resolution. This redundancy eliminates most of the numerical uncertainty so that together these four graphs possess the property that
\begin{equation}\label{eqn:planargapunion1}
\mathbb{G} = \GG(P_a) \bigcup \GG(P_b) \bigcup \GG(P_c) \bigcup \GG(P_d) \supset [-3, 2 \sqrt{2}].    
\end{equation}
Therefore any point $\xi \in [-3, 2 \sqrt{2}]$ is planar gapped by at least one of these four graphs. Supplementing these four with the existence of (necessarily non-planar) Ramanujan graphs establishes that any $\xi \in [-3,3)$ may be gapped.

Furthermore, these four examples may be used as inputs to the map $\TT$ discussed in Section \ref{sec:T}, which then transfers their gaps to other locations according to the map $f^{-1}$. Because $[-2,0] \subset \mathbb{G}$, every point in $[-3,3)$ is in the image of $\mathbb{G}$ for some power of $f^{-1}$. These graphs therefore complete the proof of Theorem \ref{thm:planargaps} and establish that every $\xi \in [-3,3)$ may be planar gapped. The only point that cannot be gapped this way is $3$. All of the Abelian covers discussed here (even the non-planar ones) have amenable deck groups and are therefore too simple to be expanders. The constant function, which has eigenvalue $3$, will always be in the closure of the $\ell^2$ space. Thus there will always be a highest band whose maximum is $3$. The action of $\TT$ cannot eliminate this band. Instead it produces gaps in the interval $[2 \sqrt{2},3)$ by compressing this band closer and closer to $3$. The study of the dynamics of $\TT$ and $f^{-1}$ in Section \ref{sec:T} shows that this band can be compressed arbitrarily, allowing for gaps underneath it and arbitrarily close to $3$.

\section{Proofs of Maximal Gap Intervals}\label{sec:maxgapproofs}

We return to the proof of Theorem \ref{thm:stronggap}
\setcounter{manualtheorem}{1}
\begin{manualtheorem}
Let $K$ be a spectral set, then 
\begin{enumerate}
\item[(i)]  $\C(K) \geq 1$.

\item[(ii)] If $I$ is an interval contained in $(-1-\sqrt{2}, 1 +\sqrt{2})$ whose length is greater than $2$, then 
$$ I \cap K \neq \emptyset.$$
\end{enumerate}
\end{manualtheorem}

For both statements in the theorem we need large geodesic segments in $X$, so we begin by producing them. If $X$ has diameter $d$ and $x_0$ and $y_0$ are in $X$ and are distance $d$ from each other, then any path $g_d$ from $x_0$ to $y_0$ of length $d$ is a goedesic (that is for any $v,w$ vertices in $g_d$ the distance from $v$ to $w$ along $g_d$ is the same as their distance in $X$). Every vertex in $X$ has distance at most $d$ from $x_0$, and hence $|X| \leq 3 \times 2^{d-1}$, the latter being the cardinality of the ball of radius $d$ in the $3$-regular tree. It follows that 
$$ d-1 \geq \log_2{\left( \frac{|X|}{3}  \right)}, $$
and hence
\begin{equation}\label{eqn:lendef}
    d > \mathbb{L}(X) := \log_2{\left( \frac{|X|}{3}  \right)}.
\end{equation}
We conclude that any $X$ contains a geodesic segment $g_d$ with $d$ satisfying Eqn. \ref{eqn:lendef}.

To prove (i) of Theorem \ref{thm:stronggap}, we must show that if $K \subset [-3,3]$ is closed and $\C (K) < 1$, then there are only finitely many $X \in \bf{X}$ with $\sigma(X) \subset K$. The spectrum $\sigma(Y)$ of the adjacency matrix of any graph $Y$ consists of numbers $\lambda$ which are real algebraic integers, all of whose conjugates are also in $\sigma(Y)$. According to Fekete's theorem \cite{Fe}, since $\C(K) <1$, there are only finitely many such algebraic integers $\lambda$ all of whose conjugates lie in $K$. It follows that the eigenvalues of any $X$ with $\sigma(X) \subset K$ must lie in a finite set $F_K$. Since $\ad_X$ is a diagonalizable, the minimal polynomial of $\ad_X$ must divide
\begin{equation}\label{eqn:polydef}
P_K(x) := \prod_{\lambda \in F_K}{(x- \lambda)} = x^k + a_{k-1}x_{k-1} + \cdots + a_0,
\end{equation}
where $k = |F_K|$.

If follows that $\ad_X$ satisfies
\begin{equation}\label{eqn:matpoly}
P(\ad_X) = \ad_X^k + a_{k-1} \ad_X^{k-1} + \cdots + a_0 \bf{I} = 0.
\end{equation}
We show that if the diameter $d(X)$ is greater than or equal to $k$, which is the case if $\mathbb{L}(X) \geq k$, then Eqn. \ref{eqn:matpoly} cannot hold. For $x, y \in V(X)$, the $x,y$ entry of the matrix $(\ad_X)^m$ in the standard basis for the adjacency matrix is equal to the number of paths in $X$ from $x$ to $y$ of length $m$. Take $x = x_0$ and $y = y_0$, where the distance from $x_0$ to $y_0$ is $k$, which can be done because $d(X) \geq k$. For $0 \leq m \leq k-1$ the $x_0,y_0$ entry of $(\ad_X)^m$ is $0$, while the entry for $(\ad_X)^k$ is not zero. Hence, the $x_0, y_0$ entry of $P(\ad_X)$ is not zero and this contradicts Eqn. \ref{eqn:matpoly}. This shows that if $\mathbb{L}(X) \geq |F_K|$, then $\sigma(X)$ cannot be contained in $K$. Thus, the set of $X$'s with $\sigma(X) \subset K$ is finite, proving Theorem \ref{thm:stronggap} (i).

To illustrate our proof of Theorem \ref{thm:stronggap} (ii), we prove a special case first. Assume that $X$ has a Hamilton path, that is a path along the edges of $X$ which passes through every vertex exactly once. Not every $X \in \bf{X}$ has such a path and, more surprisingly, nor does every planar $X$ \cite{R-W,Tu}, but most do.

\begin{proposition}\label{prop:Hampath}
If $X \in \bf{X}$ has a Hamilton path, then for $-2 \leq \lambda \leq 2$
$$ \mbox{distance} (\lambda , \sigma(X)) \leq \left( 1 + \frac{16}{|X|}   \right)^{1/2}.$$

\end{proposition}

\textit{Proof:} By assumption the vertices of $X$ can be labeled by a path $v_1, v_2, \ldots, v_t$ with $t = |X|$. Let $\ff: V(x) \rightarrow \mathbb{C}$ be given by 
\begin{equation}\label{eqn:statefw}
    \ff (v_j) = w^j, \, j = 1, \ldots, t,
\end{equation}
where $w$ is a function of $\lambda$ and satisfies
\begin{equation}\label{eqn:wquadratic}
w^2 - \lambda w + 1 = 0.
\end{equation}
Since $-2 \leq \lambda \leq 2$ we have that 
\begin{equation}\label{eqn:wsoln}
|w| = 1 \mbox{,  and  } w + w^{-1} = w + \bar{w} = \lambda  .  
\end{equation}
For $2 \leq j \leq t-1$
$$  \ad_X \ff (v_j) - \lambda \ff (v_j) = \ff (v_{j+1}) + \ff (v_{j-1}) + \ff (\hat{v}_j) - \lambda \ff(v_j),$$
where $\hat{v}_j$ is the third vertex adjacent to $v_j$. 

Hence for $2 \leq j \leq t-1$,
\begin{equation}\label{eqn:midbound}
    |\ad_X \ff (v_j) - \lambda \ff (v_j)| = 1.
\end{equation}
For $j = 1$ (and similaly for $j = t$)
$$ \ad_X \ff (v_j) - \lambda \ff (v_j) = w^2  + \ff (\hat{v}_j) + \ff (\hat{\hat{v}}_1) - \lambda w,$$
so that for $j=1$ and $j = t$
\begin{equation}\label{eqn:endbound}
|\ad_X \ff (v_j) - \lambda \ff (v_j)| \leq 3.
\end{equation}
Hence
\begin{equation}\label{eqn:normbound}
\sum_{v \in V(X)}{|\ad_X \ff (v) - \lambda \ff (v)|^2} \leq 18 + t -2 = t + 16.
\end{equation}
On the other hand
\begin{equation}\label{eqn:normbound2}
    ||\ff||^2_2 = \sum_{v \in V(X)}{|\ff(v)|^2} = t.
\end{equation}
Hence
\begin{equation}\label{eqn:normmax}
   \frac{||\ad_X \ff - \lambda \ff ||^2_2}{||\ff||^2_2} \leq 1 + \frac{16}{t}. 
\end{equation}

If the eigenvalues and corresponding orthonormal basis of eignefunctions of $X$ are denoted by $\lambda_1, \lambda_2, \cdots, \lambda_t$ and $\phi_1, \cdots, \phi_t$, then we expand $\ff$ as
$$\ff  = \sum_{j = 1}^t{\left< \ff, \phi_j \right> \phi_j}, $$
from which we find
$$ \ad_X \ff - \lambda \ff = \sum_{j = 1}^t {\left< \ff, \phi_j \right> (\lambda_j - \lambda) \phi_j}.$$
Hence
$$||\ad_X \ff - \lambda \ff ||^2_2 = \sum_{j=1}^t{(\lambda_j - \lambda)^2  \left< \ff, \phi_j \right>^2},$$
so that if $distance(\sigma(X), \lambda) := \min_j |\lambda_j - \lambda| = \beta$, then
\begin{equation}\label{eqn:normmin}
    ||\ad_X \ff -\lambda \ff||^2_2 \geq \beta^2 ||\ff||^2_2.
\end{equation}
Combining Eqns. \ref{eqn:normmax} and \ref{eqn:normmin} yields Proposition \ref{prop:Hampath}.

Our main result in this section is the following theorem which establishes Proposition \ref{prop:Hampath} for any large $X$, but with some restrictions on $\lambda$.

\begin{theorem}\label{thm:gapbound}
Let $-\sqrt{2} \leq \lambda \leq \sqrt{2}$ and $X \in \bf{X}$, then $$\mbox{distance}(\lambda, \sigma(X)) \leq \left( 1 + \frac{18}{\mathbb{L}(X)}     \right)^{1/2}.$$
\end{theorem}

\begin{figure}[h]
	\begin{center}
		\includegraphics[width=1.0\textwidth]{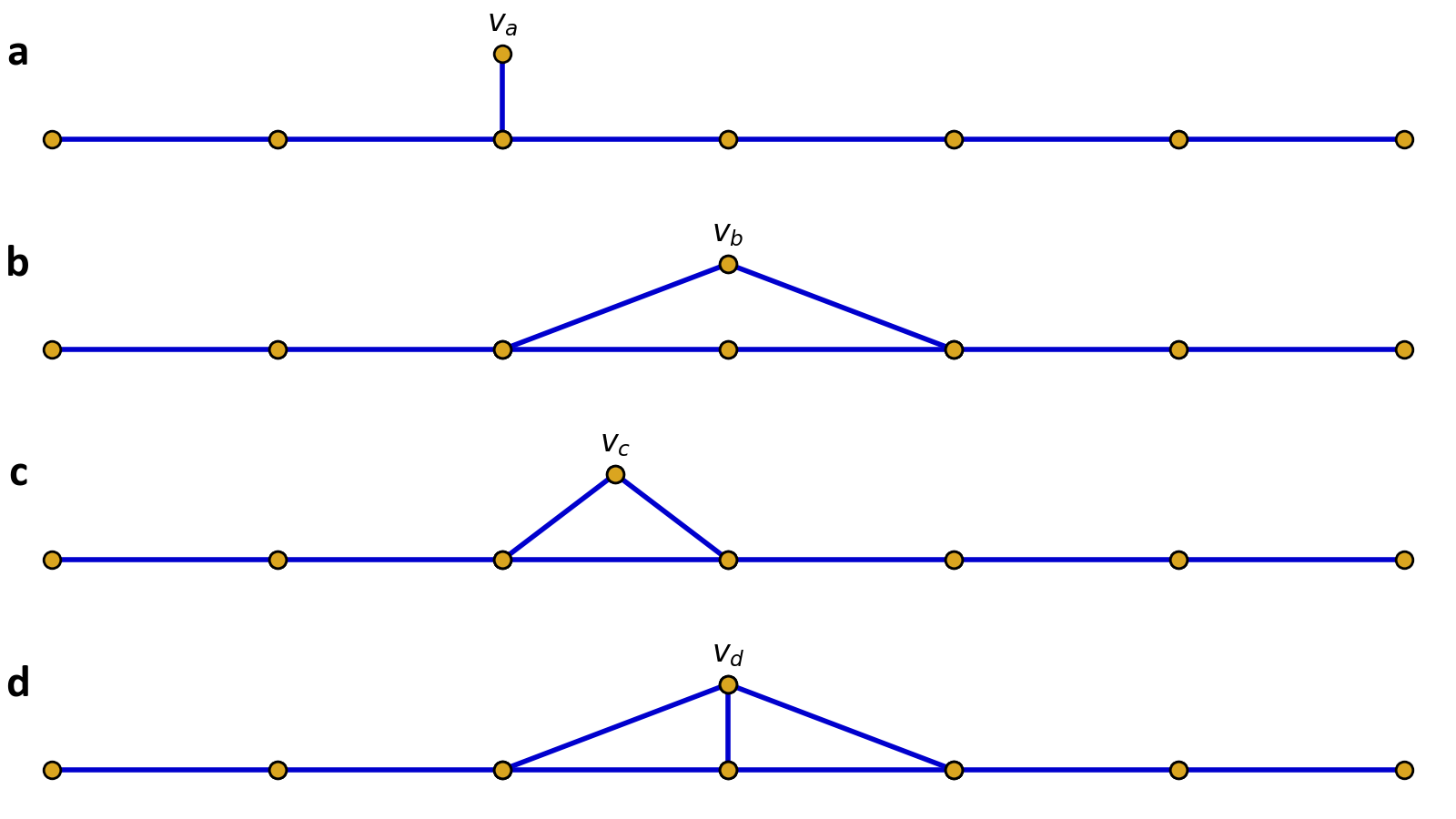}
	\end{center}
	\vspace{-0.6cm}
	\caption{\label{fig:vertTypes} 
    \textbf{Vertices connected to the geodesic.} 
    All vertices directly connected to the geodesic $g_t$ come in four types, illustrated in \textbf{a}-\textbf{d}.
    } 
\end{figure}

\begin{figure}[h]
	\begin{center}
		\includegraphics[width=1.0\textwidth]{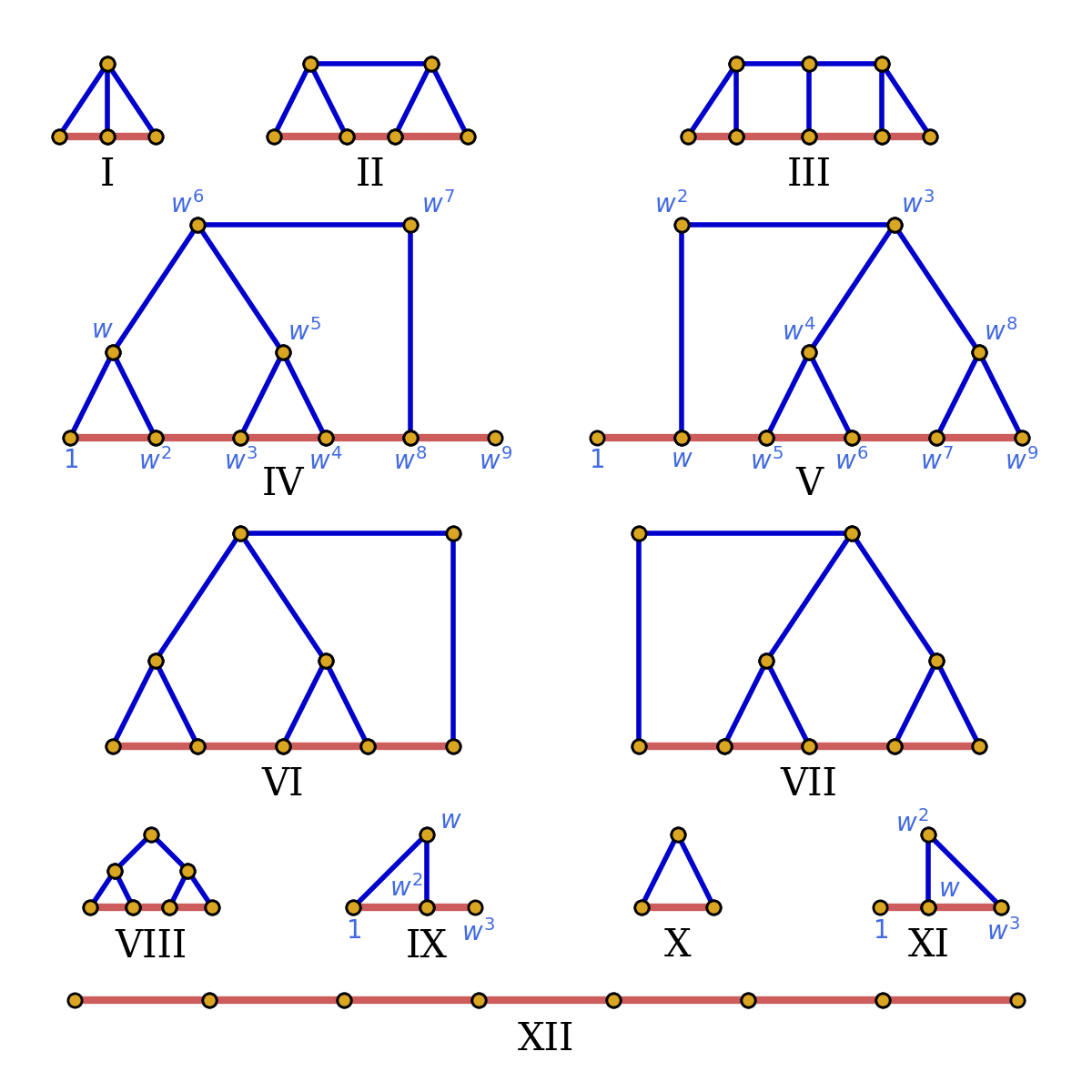}
	\end{center}
	\vspace{-0.6cm}
	\caption{\label{fig:neighborhoodTypes} 
    \textbf{Neighborhoods along the geodesic.} 
    Enumeration of all of the possible segments $S$ along the geodesic $g_t$ which contain vertices of types (c) and (d). In all cases the contained portion of the geodesic is the horizontal chain running along the bottom, indicated by  thicker, lighter-colored edges. The values of the test function $\ff$ defined along the Hamilton path of the segments is shown for types IV, V, IX, and XI.
    } 
\end{figure}

\begin{figure}[h]
	\begin{center}
		\includegraphics[width=0.95\textwidth]{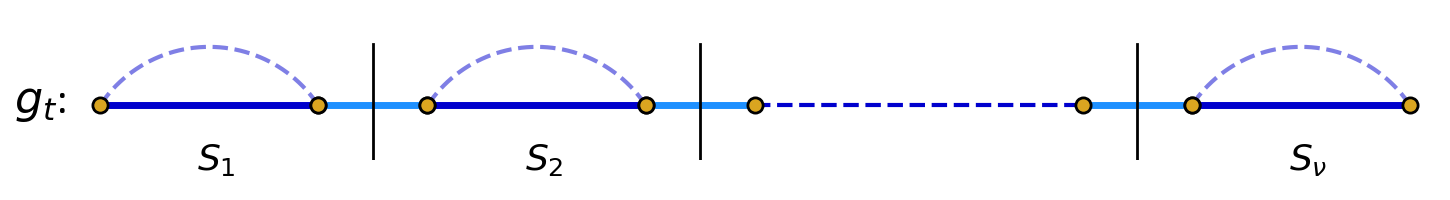}
	\end{center}
	\vspace{-0.5cm}
	\caption{\label{fig:SConnection} 
    \textbf{Segments along the geodesic.} 
    Sketch of how the geodesic $g_t$ is made up of a series of segments of the types shown in Fig. \ref{fig:neighborhoodTypes} connected only by their end points $\beta_j$ and $\alpha_{j-1}$. Connections between neighboring segments are indicated in light blue with vertical deviding lines to guide the eye.
    } 
\end{figure}

\textit{Proof:} In place of the Hamilton path that was used in Proposition \ref{prop:Hampath}, we use a long geodesic $g_t$ of length $t$ induced in $X$. According to Eqn. \ref{eqn:lendef}, such a geodesic exists from some $t > \mathbb{L}(X)$. The support of the test function $\ff : V(X) \rightarrow \mathbb{C}$ will be chosen to be a neighborhood $N_t$ of the geodesic $g_t$, and it is chosen according to how $g_t$ embeds into $X$. The vertices $v \in V(X) \backslash g_t$ which are directly (i.e. distance one) connected to $g_t$ come in four types, each shown in Fig. \ref{fig:vertTypes}.
\begin{enumerate}
    \item [$v_a$:] Connected to a single vertex in $g_t$.
    
    \item[$v_b$:] Connected to two vertices in $g_t$ which are of distance two in $g_t$.
    
    \item[$v_c$:] Connected to two vertices in $g_t$ which are neighbors in $g_t$.
    
    \item[$v_d$:] Connected to three consecutive vertices of $g_t$.
\end{enumerate}
That these are the only possibilities follows from $g_t$ being a geodesic, a feature that will be used repeatedly to limit the possible configurations, shown in Fig. \ref{fig:neighborhoodTypes}. Using the occurrences of vertices of types $c$ and $d$, we define a neighborhood $N_t$ of $g_t$ using the list in Fig. \ref{fig:neighborhoodTypes}. In all cases $S$ of the figure, the bottom horizontal segment is part of the geodesic $g_t$. Let $\alpha_s$ denote the left end vertex of this geodesic part of $S$ and $\beta_s$ the right end vertex. We claim that we can decompose $g_t$ into segments $S_j, \, 1 \leq j \leq \nu$, such that the segments link together in a chain and contain the entire geodesic $g_t$, using only segments of the types shown in Fig. \ref{fig:neighborhoodTypes}. A sketch of such a decomposition is shown in Fig. \ref{fig:SConnection}. In this form, $\beta_{S_{j}}$ is connected to $\alpha_{S_{j+1}}$ along $g_t$. (Note that a segment of type XII can be of any length.) The point of this decomposition is that only type (a) and type (b) vertices remain joined to the segments of type XII, as the other types are accounted for by the types I to XI.

To see that this can be done we go over $g_t$ looking for segments supporting I to XI from top to bottom of the table. For example, if we find a type I segment (i.e. a type (d) vertex), then it defines one such $S_j$, since the only places that a neighborhood like I can continue in $g_t$ are at $\alpha$ and $\beta$ (since the other vertices have degree $3$). By moving down the table and using the fact that $g_t$ is a geodesic, one checks that the $S_j$'s can be chosen so that there are no vertices of $V(X) \backslash g_t$ which are joined directly to different $S_j$'s. In other words, if $N_t$ is the graph consisting of the $S_j$'s $1 \leq j \leq \nu$ connected along $g_t$ as above, then
\begin{enumerate}
    \item [(1)] For any $S_j$ which is of type XII, the $v$'s not in $N_t$ joined to $S_j$ are of type (a) or (b).
    
    \item[(2)] Any $v \in V(X) \backslash N_t$ which is directly joined to some $S_j$ is not directly joined to another $S_k, \, k \neq j$.
\end{enumerate}

The $S$'s in Fig. \ref{fig:neighborhoodTypes} all have a Hamilton path running from $\alpha_S$ to $\beta_S$. These are indicated for cases IV, V, IX, and XI. In this way we obtain a Hamilton path on $N_t$ starting from $\alpha_{S_1}$ and ending at $\beta_{S_\nu}$. First traverse $S_1$ from $\alpha_{S_1}$ to $\beta_{S_1}$ using the $S_1$ Hamilton path, then cross to $\alpha_{S_2}$ (uniquely) via $g_t$, and continue. Using this labeling set $\ff : V(X) \rightarrow \mathbb{C}$ to be
\begin{equation}\label{eqn:genfunc}
\ff(v)= \begin{cases} w^k \mbox{     for } v_k \mbox{ the } k^{th} \mbox{ vertex in } N_t\\
		 0 \mbox{    if } v \notin N_t.
		  \end{cases} 
\end{equation}

We  turn to estimating
\begin{equation}\label{eqn:gennorm}
    R(\ff) = \frac{||\ad_X \ff - \lambda \ff||^2_2}{||\ff||^2_2}
\end{equation}
in order to apply En. \ref{eqn:normmin}. Clearly 
\begin{equation}\label{eqn:genbound}
||\ff||^2_2 = |N_t| \geq t \geq \mathbb{L}(X).
\end{equation}
We estimate the contributions to the numerator in Eqn. \ref{eqn:gennorm} coming from each $S_j$ separately.

If $S$ is not of type XII or IV, V, IX, XI, and we assume that $S$ does not contain one of the two end points of $g_t$, then for $v\in V(S)$
\begin{equation}
|\ad_X \ff(v) - \lambda \ff(v)| = \begin{cases} 1 \mbox{ if the degree of } v \mbox{ in } S \mbox{ is } 3 \mbox{ or if } v = \alpha_S \mbox{ or } \beta_S \\
		 0 \mbox{ otherwise, since deg} (v) \mbox{ in } S \mbox{ is } 2.
		  \end{cases} 
\end{equation}
The unique $\hat{v} \in V(S)$ which gives $0$ is connected to a vertex in $V(X)$ that is not directly connected to any $v\in V(N_t)$ other than $\hat{v}$ itself (by statement (2) above).
Hence
\begin{equation}\label{eqn:genboundS}
    \sum_{\substack{v \in V(S), \\ v \notin V(N_t) \\ \mathsf{ connected \, to \, } S  }}{|\ad_X \ff(v) - \lambda \ff(v)|^2} = |S| - 1 +1 = |S|.
\end{equation}

\begin{figure}[h]
	\begin{center}
		\includegraphics[width=0.9\textwidth]{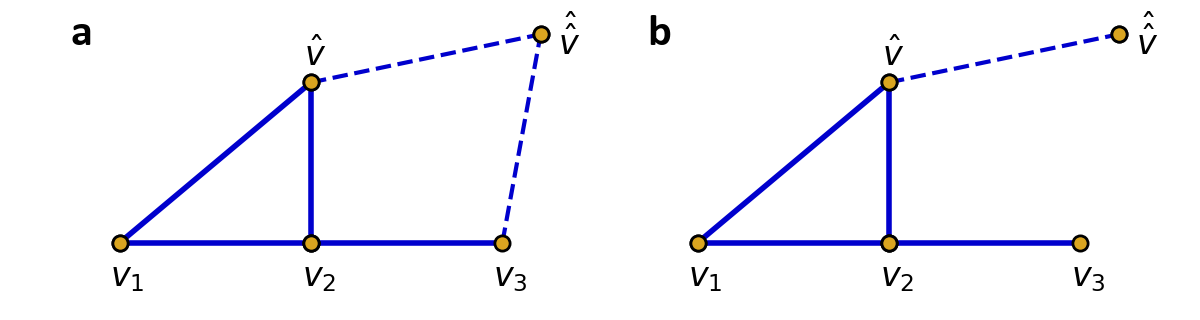}
	\end{center}
	\vspace{-0.6cm}
	\caption{\label{fig:hatHat} 
    \textbf{Second neighbors of a type IX segment.} 
    Every type IX segment has a single vertex $\hat{v}$ which is not part of the geodesic $g_t$ and is of degree two. This vertex is connected to a single additional vertex $\hat{\hat{v}} \in X$. In some cases (\textbf{a}) $\hat{\hat{v}}$ is also connected to $g_t$ and in other cases (\textbf{b}) it is not. Similar situations occur for segments of types IV, V, IX, XI.
    } 
\end{figure}

For $S$ of type IV, V, IX, XI, the analysis is a little bit different since the degree two vertex $\hat{v}$ may have a $\hat{\hat{v}}$ joined to itself and also $\alpha_S$ or $\beta_S$. For example, with type IX, we might have either of the two configurations shown in Fig. \ref{fig:hatHat}. For these configurations
$$ \ad_X \ff (\hat{v}) = \ad_X \ff (v_3) = 0, \ \ |\ad_X\ff (v_1)| = |\ad_X \ff (v_2) | = 1$$
$$ \mbox{ and } |\ad_X \ff (\hat{\hat{v}})| = |w^2 +1 | = |\lambda| $$
in the first case, 
$$ \mbox{ and } |\ad_X \ff (\hat{\hat{v}})| = |1|$$
in the second. Hence,
\begin{equation}\label{eqn:genboundS2}
\sum_{\substack{v \in V(S) \\ \mathsf{ and }\, v = \Hat{\Hat{v}}}}{|\ad_X \ff (v) - \lambda \ff (v)|^2 \leq 2  + \max{(1, |\lambda|^2)}}.
\end{equation}
Since we have assumed that $|\lambda| \leq \sqrt{2}$, we conclude that for this $S$ (and the same applies to $S$ of type IV, V, XI) that 
\begin{equation}\label{eqn:genboundS3}
  \sum_{\substack{v \in V(S) \\  v = \Hat{\Hat{v}}}} { |\ad_X \ff (v) - \lambda \ff (v)|^2 } \leq |S|.
\end{equation}
Thus, Eqn. \ref{eqn:genboundS} holds for all $S$ not of type XII.

\begin{figure}[h]
	\begin{center}
		\includegraphics[width=1.0\textwidth]{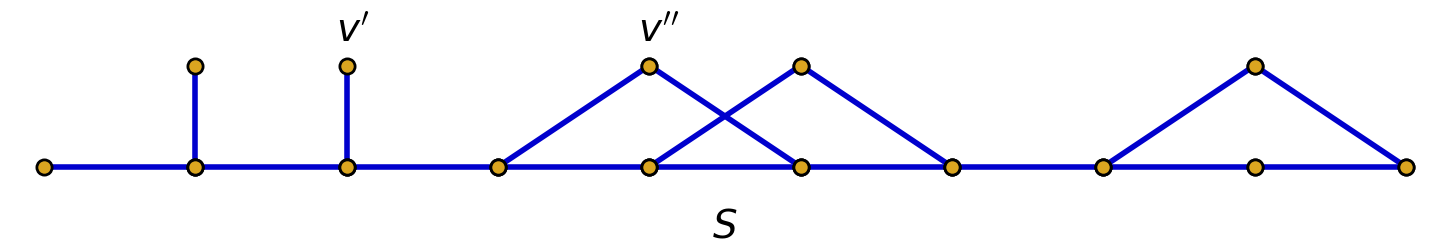}
	\end{center}
	\vspace{-0.3cm}
	\caption{\label{fig:lineTypeNeighbors} 
    \textbf{Neighboring vertices of type-XII segments.} 
    Sketch of a segment of type XII and as set of possible directly connected vertices. One type of neighboring vertex is denoted by $v'$, and is of type (a). The other type of vertex, denoted by $v''$ is of type (b). No other types are possible.
    } 
\end{figure}

Finally, for $S$ of the last type, $S$ is a geodesic segment of size $|S|$ and all $v$'s in $V(X) \backslash S$ which are joined to $S$ are of type (a) or type (b), an example of which is shown in Fig. \ref{fig:lineTypeNeighbors}.
Hence, for $v \in S$
$$ (\ad_X \ff - \lambda \ff )(v) = 0,$$
while for $v'$ of type (a)
$$|(\ad_X \ff - \lambda \ff )(v')| = 1, $$
and for $v''$ of type (b)
$$ |(\ad_X \ff - \lambda \ff )(v'')| = |\lambda|.$$
Since we have taken $|\lambda| \leq \sqrt{2}$, we conclude that 
\begin{equation}\label{eqn:genboundS4}
\sum_{\substack{v \in V(S), \\ v \, \mathsf{ directly } \\ \mathsf{joined \, to }\, S}}{|\ad_X \ff (v) - \lambda \ff (v)|^2} \leq |S| 
\end{equation}
If we add the contributions above, we get
\begin{equation}\label{eqn:genboundS5}
\sum_{j = 1}^\nu{\sum_{\substack{v \in V(S), \\ v \mathsf{\, directly } \\ \mathsf{joined \, to\,  }S}}{|\ad_X \ff (v) - \lambda \ff (v)|^2}}  \leq \sum_{j=1}^\nu{|S_j|} = |N_t|.
\end{equation}

The above assumes that the $v$'s in $g_t$ that were encountered were not one of the two extreme end points of $g_t$. For those one can get a contribution of at most $3$. Hence,
\begin{equation}\label{eqn:genboundfinal}
\sum_{v \in V(X)}{|\ad_X \ff (v) - \lambda \ff (v)|^2} \leq |N_t| + 18.
\end{equation}
Hence,
\begin{equation}
R(\ff) \leq 1 + \frac{18}{N_t} \leq 1 + \frac{18}{\mathbb{L}(X)},
\end{equation}
which, together with Eqn. \ref{eqn:normmin}, completes the proof of Theorem \ref{thm:gapbound}.

An immediate consequence of Theorem \ref{thm:gapbound} is that if $I \subset (-1-\sqrt{2} , 1 + \sqrt{2})$ and has length larger than $2$, then $I \cap \sigma(X) \neq \emptyset$ for $X$ large, which proves part (ii) of Theorem \ref{thm:stronggap}. Indeed, if $\lambda$ is the midpoint of $I$, then 
$\lambda \in [-\sqrt{2} , \sqrt{2}]$, and $(\lambda-\delta, \lambda +\delta) \subset I$ for some $\delta > 1$. Theorem \ref{thm:gapbound} then implies that $\sigma(X) \cap (\lambda - \delta, \lambda + \delta)$ is non-empty for $X$ large enough, and thus $\sigma(X) \cap I$ is also non-empty.

The above applies to any interval $I \supsetneqq (-1,1)$ or $\supsetneqq (-2,0)$. Combining this with the the constructions in Section \ref{sec:examples} showing that these intervals are achievable gap intervals leads to the conclusion that they are also maximal gap intervals. This completes the proof of Theorem \ref{thm:extremalgaps}.

To end this section, we remark that the method used to prove Theorem \ref{thm:gapbound} can be extended to cover the range $-2 \leq \lambda \leq 2$, showing that Theorem \ref{thm:stronggap} (ii) holds for any interval of length bigger than $2$. To do so requires extending the neighborhood $N_t$ further to account for the vertices of type (b). The list of special segments corresponding to Fig. \ref{fig:neighborhoodTypes} grows substantially, and since we have no immediate application of this extension, we omit the proof.

\section{Conclusion}\label{sec:conclusion}

To conclude, we elaborate on the entries in Tables \ref{table:gapints} and \ref{table:specsets} as well as some related extremal spectral sets.

The maximal gap interval $(2\sqrt{2}, 3)$ is, as noted in the Introduction, the Alon-Boppana interval \cite{Ni}. Until recently the only known construction of $Y$'s avoiding this interval was using number-theoretic tools, specifically proven cases of the Ramanujan conjectures \cite{L-P-S,Ma}. A construction of such $Y$'s using techniques from interlacing polynomials and variants of the Lee-Yang theorem \cite{H-L} was achieved in \cite{M-S-S}. That $Y$'s achieving the gap cannot be planar (in fact any sequence of $Y$'s which is $(3-\epsilon, 3)$-gapped with $\epsilon > 0$ cannot be planar) follows from the separator theorem \cite{L-T}. 

The ``Hoffman interval'' $[-3,-2)$ has been discussed and exploited repeatedly throughout the paper and especially its characterization in terms of the map $\TT$ (Section \ref{sec:T} (\ref{eqn:Tpropsv})). The gap intervals $(-2,0)$ and $(-1,1)$ are analyzed in Sections \ref{sec:examples} and  \ref{sec:maxgapproofs}.

In the context of planar graphs, gaps at points other than $3$ and $-3$ are important in a variety of contexts. The gap between the smallest of the upper half of the eigenvalues and the largest of the lower half is a measure of the Huckel stability of carbon Fullerenes \cite{F-M, Fo, M-W-F}. 
It is also decisive in the properties of materials such as carbon, where electrons fill half of the available states.
Barring non-linear effects, lattices with a gap at this point in the spectrum are insulating, and those without are conducting \cite{Gi}. Other chemical compositions or doping levels will lead to other relevant fractions. The size of this gap is also critical in distinguishing, e.g., the semiconductors that power modern electronics with relatively small gaps from strongly insulating materials with much larger ones.

We have shown that gaps can be created for planar cubic graphs; however, if one limits the types of faces in such graphs, then it is much more difficult to produce gaps. We examine this phenomenon in forthcoming joint work with Fan Wei, where we show that $(-1,1)$ is the unique maximal gap set for planar cubic graphs which have at most six sides per face. On the other hand, every point in $[-3,3)$ can be gapped for planar graphs with at most $64$ sides per face. For Fullerenes, that is planar cubic graphs with twelve pentagon faces and the rest hexagons, the only points that can be gapped are those in $(-E,E) \backslash \{ \lambda_b\}$, where
$$ E = \sqrt{1 + 4 \cos(\pi/10) \cos (21\pi/30) + 4 \cos^2 (21 \pi/30)}  = 0.382\ldots,$$
and 
\begin{eqnarray}
\lambda_b & = & 0.360\ldots \nonumber\\
 &=&   \frac{\left[\tubeFsquared + \left( \tubeF-1 \right) ^2   \right]}{2 \left(\tubeF-1 \right) }  - \nonumber
\end{eqnarray}
\begin{equation}\nonumber
\frac{\sqrt{\left[\tubeFsquared + \left(\tubeF-1 \right) ^2  \right] ^2 + 4 \left[ \tubeF-1\right] ^2\left[ 1-\tubeFsquared \right]    }}{2(\tubeF-1)}. \nonumber
\end{equation}
In particular, for any sequence of leapfrog Fullerenes \cite{M-W-F}, no point in $[-3,3]$ can be gapped, which answers the question of whether such gaps can exists which was raise in (Discussion of Figure 1(a)) in Ref. \cite{Fo}.

Another question about the gap interval $(-1,1)$ that we do not know the answer to is whether it is a maximal gap \textit{set}. $(-2,0)$ is definitely not a maximal gap set since, unlike the $(-1,1)$ case, the $Y$'s in Fig. \ref{fig:extremals} \textbf{a}\textit{ii} have a small gap below $2$ as well.

We turn to the minimal spectral set $K_1 = [-2 \sqrt{2} , 2 \sqrt{2}] \cup \{3 \}$. The fact that $K_1$ is spectral follows from the construction of non-bipartite Ramanujan graphs, which to date have only been achieved using number theory. That $K_1$ is minimal follows from \cite{A-G-V}, who show that any growing sequence $Y_m$ of non-bipartite (cubic) Ramanujan graphs Benjamini-Schramm converges to the $3$-regular tree. This in turn implies that the density of states probability measures
$$ \mu(Y_m) := \frac{1}{|Y_m|}   \sum_{\lambda \in \sigma(Y_m)}{ \delta_\lambda},$$
$\delta_\lambda$ being a point mass at $\lambda$, converge to the adjacency density for the $3$-regular tree. (Under the assumption that the girths of the $Y_m$'s go to infinity this was shown in \cite{Se2}, p100.)
The latter was computed by Kesten \cite{Ke} and its support is $[-2 \sqrt{2}, 2 \sqrt{2}]$, and hence $K_1$ is minimal.

According to Proposition \ref{prop:Kspectral}, $K_2 = f^{-1}(K_1) \cup \{ 0,-2\}$ is a minimal spectral set. Repeating this yields an infinite sequence $K_n$ of minimal spectral sets which interpolate between the fattest, $K_1$, and the thinnest, $A$, such sets.

We showed that $A$ has the smallest capacity among these sets, and we conjecture that $K_1$ has the largest, namely $\sqrt{2}$. Note that if $(-1,1)$ is a maximal gap set, then $[-3,3] \backslash (-1,1)$ would be another minimal spectral set with capacity equal to $\sqrt{2}$ \cite{G-V} (Apply Theorem 11 with $E_0 = [-4,4]$ and $f(z) = z^2 -5$).
Finally, given that every $\xi \in [-3,3)$ is planar gapped, it would be interesting to work towards a description of realizable gap sets and investigate the sizes of the maximal gap intervals about $\xi$ for various $\xi$. 
The spectral gap questions that we have investigated here for cubic graphs can be posed more generally for $r$-regular graphs ($r > 3$). Many of the techniques that we have used apply to these and it would be interesting to pursue such a study.


\begin{ack}
We would like to thank N. Alon, A. Chapman, A. Gamburd, J. Koll\'{a}r, P. Kuchment, N. Linial, B. Mohar, S. Flammia, and  F. Wei for instructive discussions related to this work.
\end{ack}

\bibliographystyle{amsplain}
\bibliography{GapSets.bib}

\end{document}